\documentclass[aps,prb,twocolumn,amssymb,amsmath]{revtex4}
\usepackage{graphicx}
\usepackage{color}
\usepackage{bm}

\newcommand*{\grad}{\mbox{\boldmath$\nabla$}}
\newcommand*{\cNf}{{\cal N}_F}
\newcommand*{\vvf}{\vec{v}_F}
\newcommand*{\vfgrad}{\vvf\cdot\grad}
\newcommand*{\vpf}{s}
\newcommand*{\vpfp}{\vpf^{\prime}}
\newcommand*{\vpfpp}{\vpf^{\prime\prime}}
\newcommand{\gvec}[1]{\mbox{\boldmath $#1$}}

\begin{document}

\title{Phase Diagrams of Ferromagnet-Superconductor Multilayers with
Misaligned Exchange Fields}

\author{Tomas L\"{o}fwander, Thierry Champel, and Matthias Eschrig}

\affiliation{Institut f\"{u}r Theoretische Festk\"{o}rperphysik,
Universit\"{a}t Karlsruhe, D-76128 Karlsruhe, Germany}

\date{\today}

\begin{abstract}
We study the influence of misalignment of the ferromagnetic exchange
field on the equilibrium properties of hybrid structures, composed of
superconducting (S) and ferromagnetic (F) parts.  In particular, we
study numerically the superconducting critical temperature $T_{c}$ in
F-S-F trilayers and in F-S-F-S-F Josephson junctions as a function of
the misalignment angle $\theta $ of the ferromagnetic magnetization.
We discuss the corresponding phase diagrams for these hybrid
structures.  For the Josephson junctions, a transition between the
zero-phase and the $\pi$-phase ground state as a function of $\theta $
takes place under certain conditions.  Within the quasiclassical
Green's function technique in the diffusive limit, we introduce a fast
and effective method for calculating $T_c$ in such multilayer
structures.
\end{abstract}

\pacs{74.45.+c,74.62.-c,74.78.Fk,74.25.Dw}

\maketitle

\section{Introduction}

The interest in superconductor-ferromagnet (S-F) hybrid structures has
considerably increased in the last decade due to their relevance for
the development of nanometer scale electronic devices.  The
understanding of the superconducting proximity effect in S-F devices
is of vital importance for such a goal. Consequently, experimental and
theoretical studies have focused on the influence that proximity
induced spin-triplet pairing amplitudes in S-F hybrid structures have
on superconducting properties of the entire structure.  Among those
are for example changes of the superconducting transition temperature
$T_c$ of the device, or the switching between 0-junctions and
$\pi$-junctions as ground states in S-F-S Josephson devices as a
function of some control parameter.

The superconducting critical temperature $T_{c}$ in diffusive hybrid
S-F structures has been studied both
theoretically\cite{Buz1990,Rad1991,Buz1992,Dem1997,Tag1999,Buz1999,Bal2001,Fom2002,Bal2003,Fom2003,You2004}
and
experimentally\cite{Jia1995,Muh1996,Aar1997,Laz2000,Gu2002,Obi2005,Cir2005,Pot2005,Mor2006,Rus2006}
in several recent publications. It has been shown
\cite{Buz1990,Rad1991} that $T_{c}$ has a non-monotonic dependence on
the thickness $d_{f}$ of the ferromagnetic layers that provide
information about the strength of the ferromagnetic exchange field and
about the transparencies of the S-F interfaces.  Approximate analytic
formulas for $T_c$ have been derived for several limiting
cases,\cite{Buz} e.g. for thin or thick film thicknesses or for low or
high interface resistances. Recently, Fominov {\it et al.} developed a
numerical method to compute $T_c$ for diffusive S-F
bilayers\cite{Fom2002} and symmetric F-S-F trilayers \cite{Fom2003}
for arbitrary model parameters such as layer thicknesses and interface
resistances. Such an approach is valuable when theory and experiments
are compared in detail with the aim to extract parameters as e.g. the
ferromagnetic exchange field or the boundary transparencies.

The possibility\cite{Buz1982,Buz1990,Buz} of a $\pi$-state
(characterized by a stable phase difference of $\pi$ between the
superconducting order parameters) is now well established
experimentally in S-F hybrid structures involving several
superconducting layers.  Transitions between the 0 and $\pi$ states
have been revealed in S-F-S junctions by the oscillations of the
critical current when varying the
temperature\cite{Rya2001,Sel2004,Fro2004} or the ferromagnetic
thickness.\cite{Kon2002,Blu2002,Gui2003} The transitions from the $0$
to $\pi$ state may also be revealed\cite{Rad1991} by the presence of
cusps in the dependence of $T_{c}$ on $d_{f}$. Because the cusps may
be confounded with the oscillations of $T_{c}(d_{f})$ themselves, such
a feature in the dependence of $T_{c}$ has been identified
experimentally only recently.\cite{Obi2005,Shelukin}

The presence of several ferromagnetic layers introduces a new degree
of freedom, the relative orientation angle, $\theta $, between the
magnetizations.  The influence of the orientation on $T_{c}$ has been
first studied theoretically in F-S-F trilayers in the
Refs.~\onlinecite{Tag1999} and \onlinecite{Buz1999} (these authors
only considered parallel or antiparallel orientations). The
calculations for an arbitrary orientation were performed in
Ref.~\onlinecite{Fom2003}. A dependence of the critical current
oscillations on the magnetization orientation has been also
established theoretically in S-F-F'-S
junctions\cite{Ber2001c,Ber2001b,Gol2002, Bla2004} and multilayered
S-F junctions.\cite{Ber2003} In Ref.~\onlinecite{Gol2002} and
\onlinecite{Ber2003} a switch between the 0 and $\pi$ states has been
found from calculations of the Josephson critical current by changing
the mutual orientation between the moments. The dependence of $T_{c}$
on the moment orientation (parallel or antiparallel) of trilayers has
been studied experimentally in
Refs.~\onlinecite{Gu2002,Pot2005,Mor2006,Rus2006}. A dependence on the
domain state of the ferromagnet in a S-F bilayer was found in
Ref.~\onlinecite{Rus2004}.

Motivated by the recent experimental studies, we have developed a fast
and effective method that is particularly suited for the numerical
calculation of $T_c$ in diffusive hybrid structures. An important part
of this paper is to present details of this method and discuss the
calculations leading to some of the results presented in
Ref.~\onlinecite{Shelukin}. Our method can be considered as a
development of the method of Fominov {\em et al.}, who in
Refs.\onlinecite{Fom2002,Fom2003} have presented calculations of $T_c$
of S-F bilayers and symmetric F-S-F trilayers with non-collinear
magnetizations. We extend the calculations to the more general case of
asymmetric trilayers in connection with the geometry considered
typically in experiments.  Within our model, we also treat symmetric
pentalayers, including the possibility of a phase difference $\pi$
between the two superconductors. This structure was recently studied
experimentally in Ref.~\onlinecite{Shelukin}. From our $T_{c}$
calculations, we predict a switching between $0$- and $\pi$-junctions
by the orientation of the ferromagnetic exchange fields in pentalayers
consisting of a central Josephson junction, two superconductors
separated by a ferromagnet, sandwiched between two outer ferromagnets
with exchange fields rotated relative to the central ferromagnetic
layer. This kind of structure could be realized e.g. by fixing the
moments of the outer layers, while rotating the moment of the central
layer with an external magnetic field.

The outline of the paper is as follows. In Section
\ref{Sec_ModelMethod} we present the model of the F-S-F trilayer and
the F-S-F-S-F pentalayer structures and outline the method that we use
to compute the order parameter profile and $T_c$ of the structures. In
Sections \ref{Sec_ResultsTrilayer} and \ref{Sec_ResultsPentalayer} we
present the results for the trilayer and pentalayer respectively. In
Section \ref{Sec_Numerics} we discuss some details of our numerical
method. We summarize our work in Section \ref{Sec_Summary}. Some of
the technical details have been collected in the appendices.

\section{Model and Method}\label{Sec_ModelMethod}

We shall restrict our considerations to diffusive hybrid structures
and to temperatures $T$ near the critical temperature $T_c$. We employ
a Green's function method in the quasiclassical approximation.  The
central quantity in this framework is the $2\times 2$ spin-matrix
anomalous Green's function $f$, describing superconducting
correlations in the structure.  The spin degree of freedom has to be
kept due to the fact that the ferromagnets in proximity with the
superconductors break spin rotational invariance. Thus, both spin
singlet and spin triplet proximity pair amplitudes are present in the
ferromagnet.  We use a notation where the spin structure is described
as $f=(f_s+{\gvec{\sigma}}\cdot{\bf f}_t)i\sigma_y$, where
${\gvec{\sigma}}=(\sigma_x,\sigma_y,\sigma_z)$ are the three Pauli
matrices. The pair amplitudes are the spin singlet component $f_s$ and
the three spin triplet components described by the vector ${\bf
f}_t$. The ferromagnetic regions are characterized by an exchange
field with a fixed direction.  In the case of rapid changes on the
scale of the coherence length of the direction of the exchange
field,\cite{Ber2001,Kad2001,Vol2003} or spin-active interface
scattering,\cite{escPRL03} long-range equal-spin triplet correlations
are also induced.  We refer to our recent
papers\cite{escPRL03,Cha2005,Cha2005b,Lof2005} and a recent
review\cite{Ber2005} and references therein for a deeper discussion of
the origin of these correlations.

For diffusive structures the Green's function is isotropic to lowest
order in $1/p_f\ell$, where $p_f$ is the Fermi momentum and $\ell$ is
the mean free path. Furthermore, for temperatures near the critical
temperature the superconducting gap is small $\Delta\ll T_c$ and the
usual Green's function is approximately equal to the normal state
Green's function $g\approx -i\pi\mbox{sgn}(\varepsilon_n)$, while the
anomalous Green's function $f$ is small, of the order of $\Delta$. The
relevant starting point in this case is Usadel's diffusion
equation\cite{Ref_Usadel} linearized for small $\Delta$. We assume for
simplicity that the spatial dependence in the structure is only along
the interface normal, taken to be along the $x$-axis, see
Figs.~\ref{Fig_tri} and \ref{Fig_penta}. Then, the linearized Usadel
equations have the form\cite{Cha2005b}
\begin{eqnarray}
\left(D \partial_{xx}^{2} -2 |\varepsilon_{n}|\right)f_{s} &= & 
- 2 \pi \Delta +2 i\, \mbox{sgn}(\varepsilon_n)\, {\bf J} \cdot {\bf f}_{t} \label{coup1}\\
\left(D \partial_{xx}^{2} -2 |\varepsilon_{n}|\right) {\bf f}_{t} &= & 
2 i\, \mbox{sgn}(\varepsilon_n)\, {\bf J} f_{s}, \label{coup2}
\end{eqnarray}
where $\mbox{sgn}(\varepsilon_n)$ is the sign of the Matsubara
frequency $\varepsilon_{n}=\pi T(2n+1)$, and we have used the short
hand notation $f=f(\varepsilon_n,x)$. We assume that the exchange
field ${\bf J}={\bf J}(x)$ is non-zero in the ferromagnetic regions,
while $\Delta=\Delta(x)$ is non-zero in the superconducting
regions. Each layer in the structure can have a different diffusion
constant $D$. Note that we assume that the exchange field is small
compared to the Fermi energy, $J\ll\epsilon_f$, in which case the
quasiclassical theory can be straightforwardly applied. For strong
exchange fields, a separate calculation has to be made.\cite{escPRL03}

The diffusion equation is supplemented with boundary conditions at
each interface and at the outer surfaces of the structure. The
boundary condition connecting the Green's function at $x_S$ on the
superconductor side of the interface with the Green's function at
$x_F$ on the ferromagnet side of the interface is of the form first
derived by Kupriyanov and Lukichev\cite{Kup}
\begin{eqnarray}
\gamma \xi_F f'(x_F) &=& \xi_S f'(x_S),\label{bound1}\\
\gamma_b \xi_F f'(x_F) &=& \pm\left[f(x_S)-f(x_F)\right],\label{bound2}
\end{eqnarray}
where $\xi=\sqrt{D/2 \pi T_{c0}}$ is the coherence length and the
parameters $\gamma$ and $\gamma_b$ characterize the conductivity
mismatch between the two sides and the boundary resistance,
respectively. The sign in Eq.~(\ref{bound2}) is positive (negative)
for a F/S (S/F) interface [for which the superconductor occupies the
space to the right (left) of the barrier]. Note that we use the prime
as a short-hand notation for spatial derivatives at a certain point in
space, e.g.  $f'(x_S)=\left.\partial_x f(x)\right|_{x=x_S}$.  At the
outer surfaces of the structure, we require that the current trough
the boundary must vanish, i.e. $\partial_{x} f=0$.

Since the exchange field and the superconducting order parameter are
spatially separated, we see that in the superconducting region
Eqs.~(\ref{coup1})-(\ref{coup2}) are decoupled. The triplet part
(\ref{coup2}) can be solved analytically, while the singlet part
(\ref{coup1}) has a source term containing the order parameter that
satisfies the self-consistency equation
\begin{equation}
\Delta(x) \, \ln \frac{T}{T_{c0}}=T \sum_{\varepsilon_n} \left( 
f_{s}(\varepsilon_n,x) - \frac{\pi\Delta(x)}{|\varepsilon_{n}|}
\right) \label{gap}.
\end{equation}

In the ferromagnetic regions Eqs.~(\ref{coup1})-(\ref{coup2}) are
coupled but the superconducting order parameter is absent and both
equations can be solved analytically, which is described in detail in
Appendix~\ref{App_trilayer} and \ref{App_pentalayer} for the trilayer
and pentalayer cases. The presence of the ferromagnetic regions are in
the process reduced to an effective boundary condition for the
calculation of the singlet component in the superconducting region,
which we confine to $0<x<d_s$ in the present discussion, as in
Fig.~\ref{Fig_tri}. The boundary condition can in the general case be
written in the form
\begin{equation}
\left(
\begin{array}{c}
f_{s}'(0)\\
f_{s}'(d_s)
\end{array}
\right)
=k_{s}\hat{W}
\left(
\begin{array}{c}
f_{s}(0) \\
f_{s}(d_s)
\end{array}
\right), \label{bound}
\end{equation}
where $k_{s}=\sqrt{2 \varepsilon_{n}/D_{s}}$ and $\hat{W}$ is a $2
\times 2$ matrix. The non-locality of the boundary condition
(\ref{bound}), i.e. the coupling of the two interfaces at $0$ and at
$d_s$, is a result of the coupling of the singlet and triplet
anomalous Green's functions $f_s$ and ${\bf f}_t$ in the original
boundary conditions Eqs.~(\ref{bound1})-(\ref{bound2}) and the
coupling of the diffusion equations for the singlet and triplet
components in the ferromagnet by the exchange field. The matrix
$\hat{W}$ depends on the Matsubara frequency, the parameters of the
adjacent layers (thicknesses, exchange fields, diffusion constants),
and the interface parameters $\gamma$ and $\gamma_b$. The expressions
for the components of $\hat{W}$ are derived in
Appendix~\ref{App_trilayer} and \ref{App_pentalayer} for the trilayer
and pentalayer structures. The following method for calculating $T_c$
is however applicable for any matrix $\hat W$, as long as the boundary
condition for the singlet Green's function $f_s$ is of the form
(\ref{bound}).

Consider Eq.~(\ref{coup1}) in the superconducting region, i.e. for
$0<x<d_s$ where ${\bf J}=0$. By linear superposition we
have\cite{Fom2002}
\begin{eqnarray}
f_{s}(\varepsilon_n,x)=\pi \int_{0}^{d_s} \,G(\varepsilon_n,x,y) \Delta(y) \,dy, \label{fsfromdelta}
\end{eqnarray}
where the function $G(\varepsilon_n,x,y)$ is the solution of the
differential equation
\begin{equation}
\left(\frac{D_{s}}{2} \partial_{xx}^{2} - |\varepsilon_{n}|\right) G(\varepsilon_n,x,y)=- \delta(x-y), \label{G}
\end{equation}
subject to the boundary conditions (\ref{bound}) with
$f_s(\varepsilon_n,x)$ replaced by $G(\varepsilon_n,x,y)$. The
solution of Eq.~(\ref{G}) is presented in Appendix~\ref{App_G}. With
the help of the function $G$, the gap equation can be written as
\begin{equation}
\frac{2\pi T\sum_{\varepsilon_n>0} \int_{0}^{d_s} \,G(\varepsilon_n,x,y) \Delta(y) \,dy}
     {\ln\frac{T}{T_c0}+2\pi T\sum_{\varepsilon_n>0}\left(\varepsilon_n\right)^{-1}}
= \Delta(x),
\label{self}
\end{equation}
where we used that the singlet Green's function $f_s(\varepsilon_n,x)$
[and therefore also $G(\varepsilon_n,x,y)$] is an even function of
$\varepsilon_n$. We see that one way\cite{Fom2002} of solving the
problem at hand is to discretize the spatial coordinate ($x\rightarrow
x_k$, $k=1...N$) and find the critical temperature $T_c$ as the
highest temperature for which the eigenvalue of the $N\times N$ matrix
on the left hand side of Eq.~(\ref{self}) equals one. The
corresponding eigenvector gives the profile of the order parameter,
$\Delta(x_k)$.

There are several disadvantages of the method described above, all
connected with the discretization of the spatial coordinate axis. In
particular, it is cumbersome to reach acceptable numerical accuracy
when $T_c$ is computed. We shall discuss these problems in detail in
Section~\ref{Sec_Numerics}.

Because of these draw backs, we develop a Fourier series method that
avoids the discretization of the spatial coordinate. The
superconducting order parameter $\Delta(x)$ exists in the range $0 < x
< d_s$. We extend its domain of definition to the full real axis by
adding an even-parity property and $2d_{s}$ periodicity. Then,
$\Delta(x)$ can be expanded in a Fourier series
\begin{equation}
\Delta(x)=\sum_{p=0}^{\infty} \Delta_{p} \cos \left(\frac{p \pi x}{d_{s}}\right), \label{fourier}
\end{equation}
where the coefficients $\Delta_{p}$ are defined as
\begin{equation}
\Delta_{p}=\frac{2-\delta_{p0}}{d_{s}} \int_{0}^{d_{s}} \, \Delta(x) \cos\left(\frac{p \pi x}{d_{s}}\right)\,dx .
\end{equation}
We show in Appendix~\ref{App_fourier} how to obtain an analytic
expression for the singlet amplitude $f_{s}$ in terms of the Fourier
coefficients $\Delta_p$. Consequently, the gap equation can be written
in the space of Fourier coefficients as
\begin{equation}
\sum_{p=0}^{\infty} m_{lp} \Delta_{p}=0, \label{eigenp}
\end{equation}
for integer $l \geq 0$, and where $m_{lp}$ are given in
Eqs.~(\ref{mlp})-(\ref{mll}). We solve the problem at hand by
introducing a cut-off $p_c$ for the number of harmonics and find the
critical temperature $T_c$ as the highest temperature for which the
eigenvalue of the $p_c\times p_c$ matrix on the left hand side of
Eq.~(\ref{eigenp}) equals zero. The corresponding eigenvector gives
the profile of the order parameter $\Delta(x)$ through the sum in
Eq.~(\ref{fourier}).

In the following two section we use this method to compute $T_c$ for
the S-F trilayer and pentalayer structures. In
Section~\ref{Sec_Numerics} we discuss the advantages of our method
[Eq.~(\ref{eigenp})] and compare with the other method
[Eq.~(\ref{self})].

\section{Trilayer}\label{Sec_ResultsTrilayer}

\begin{figure}[t]
\includegraphics{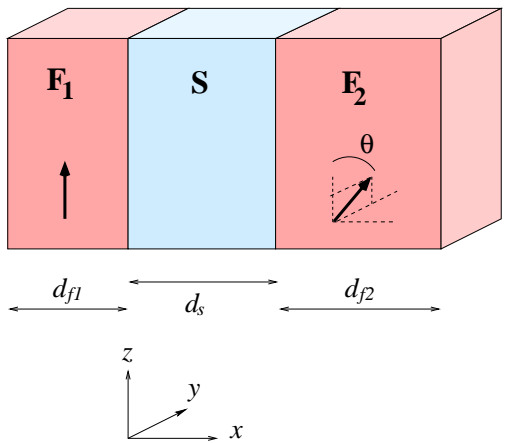}
\caption{Geometry of the asymmetric F$_{1}$-S-F$_{2}$ structure. The
moments ${\bf J}_{1}$ (in F$_{1}$) and ${\bf J}_{2}$ (in F$_{2}$) may
have different amplitudes and point in different directions (the
relative orientation angle is denoted $\theta$).}
\label{Fig_tri}
\end{figure}

Consider the trilayer structure shown in Fig.~\ref{Fig_tri}. We study
in this section the superconducting transition temperature of such a
trilayer. Our studies are motivated by the recent experiments on S-F
layered
structures,\cite{Jia1995,Muh1996,Aar1997,Laz2000,Gu2002,Obi2005,Cir2005,Pot2005,Mor2006,Rus2006}
including in particular the experiments in Ref.~\onlinecite{Shelukin}
on the critical temperature of asymmetric F$_1$-S-F$_2$ trilayers. The
theory fits of $T_c$ of the trilayers in Ref.~\onlinecite{Shelukin}
were obtained with the theory presented in the present paper.

In the left ferromagnetic layer (F$_{1}$), the exchange field is
aligned with the $z$-axis, while in F$_{2}$ it lies in the $yz$ plane
and forms an angle $\theta$ with respect to the $z$-axis. The origin
of the coordinate system is taken at the F$_{1}$/S interface. The two
layers F$_1$ and F$_2$ are characterized by their thicknesses
($d_{f1}$, $d_{f2}$), exchange fields ($J_{1}$, $J_{2}$) and diffusion
constants ($D_{f1}$, $D_{f2}$), while the superconducting layer is
characterized by its thickness ($d_s$), pairing interaction strength
(i.e. the bulk material superconducting critical temperature
$T_{c0}$), and diffusion constant ($D_s$). The diffusion constants are
converted into coherence lengths $\xi=\sqrt{D/2\pi T_{c0}}$ and we
shall use the coherence length in the superconductor $\xi_s$ as length
scale in the problem.  The F$_{1}$/S and S/F$_{2}$ interfaces are
characterized by the conductivity mismatches ($\gamma_1$, $\gamma_2$)
and interface resistances ($\gamma_{b1}$, $\gamma_{b2}$).

The Usadel equations (\ref{coup1})-(\ref{coup2}) are solved as
described in Appendix~\ref{App_trilayer} to give the effective
boundary condition matrix $\hat W$ for the trilayer. The matrix
$m_{lp}$ of Eq.~(\ref{eigenp}) is then given in terms of the elements
of $\hat W$ as shown in Appendix~\ref{App_fourier}.

\subsection{Results}

\begin{figure}[th]
\includegraphics[width=8cm]{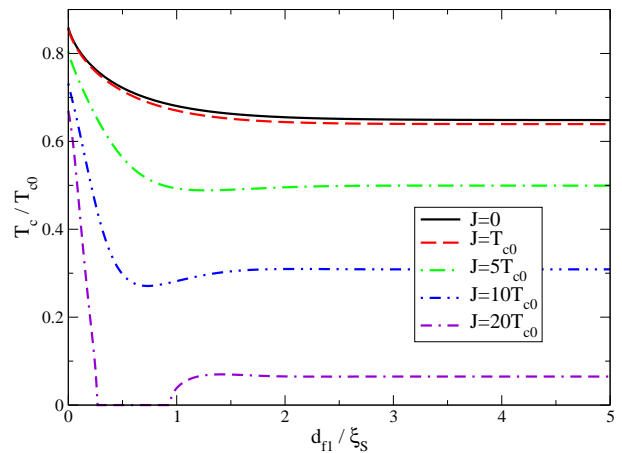}
\caption{Critical temperature $T_{c}$ of a trilayer versus the
thickness of the left ferromagnet for several strengths of the
exchange field ranging from $J_{1}=J_{2}=J=0$ (the normal metal case)
to strong $J=20T_{c0}$. The other layer thicknesses are $d_s=2\xi_S$
and $d_{f2}=0.5\xi_S$, the interface parameters are
$\gamma_1=\gamma_2=0.3$ and $\gamma_{b1}=\gamma_{b2}=0.7$, the
exchange fields are parallel ($\theta=0$), and the diffusion constants
are equal, $D_{f1}=D_{f2}=D_S$.}
\label{Fig_tri_Tc-vs-df1-and-J}
\end{figure}

\begin{figure}[th]
\includegraphics[width=8cm]{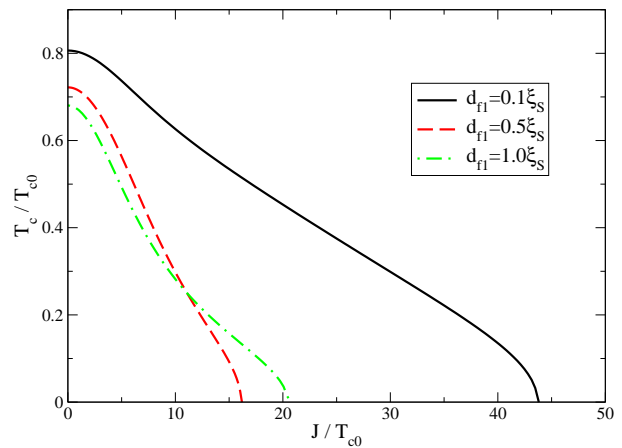}
\caption{Critical temperature $T_c$ of a trilayer versus the strength
of the exchange fields $J_1=J_2=J$ for a few layer thicknesses
$d_{f1}$ of one of the ferromagnetic layers. The other parameters are
the same as in Fig.~\ref{Fig_tri_Tc-vs-df1-and-J}.}
\label{Fig_tri_Tc-vs-J-and-df1}
\end{figure}

In Fig.~\ref{Fig_tri_Tc-vs-df1-and-J}-\ref{Fig_tri_Tc-vs-J-and-df1} we
present the influence of an exchange field on $T_c$ for an asymmetric
F$_1$-S-F$_2$ trilayer (with $d_{f1} \neq d_{f2}$). In the normal
metal case (obtained by setting $J=0$), the critical temperature is
monotonically suppressed as the layer thickness $d_{f1}$ is increased,
see Fig.~\ref{Fig_tri_Tc-vs-df1-and-J}. In the case of a ferromagnet,
the exchange field induces an additional oscillatory behavior, closely
connected to the spin mixing between up and down spins. As a result,
$T_{c}$ is suppressed in a non-monotonic way.  For a strong enough
exchange field, the oscillation is so strong that superconductivity is
suppressed at a critical thickness but can reappear at a larger
thickness. This kind of non-monotonic dependence of $T_c$ was
thoroughly
studied\cite{Dem1997,Tag1999,Buz1999,Fom2002,Fom2003,You2004,Buz} for
F-S bilayers and symmetric F-S-F trilayers. The crossing of the two
$T_{c}$ curves in Fig.~\ref{Fig_tri_Tc-vs-J-and-df1} is due to the
non-monotonic $d_{f1}$ dependence shown in
Fig.~\ref{Fig_tri_Tc-vs-df1-and-J}.

\begin{figure}[th]
\includegraphics[width=8cm]{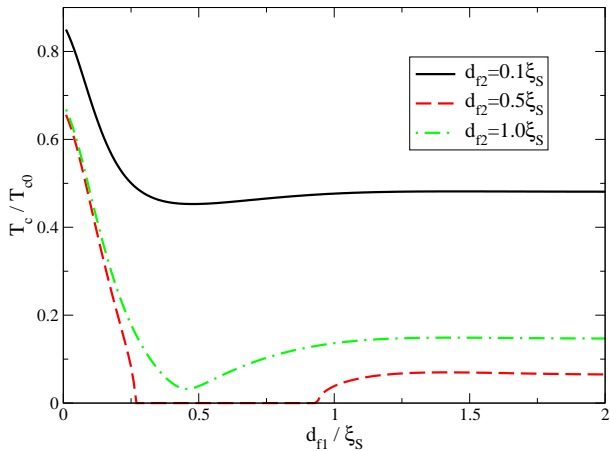}
\caption{Critical temperature $T_{c}$ of a trilayer versus the
thickness $d_{f1}$ of the left ferromagnet for several thicknesses
$d_{f2}$ of the right ferromagnet. The exchange field is
$J_1=J_2=20T_{c0}$ and the other parameters are the same as in
Fig.~\ref{Fig_tri_Tc-vs-df1-and-J}.}
\label{Fig_tri_Tc-vs-df1-and-df2}
\end{figure}

\begin{figure}[th]
\includegraphics[width=8cm]{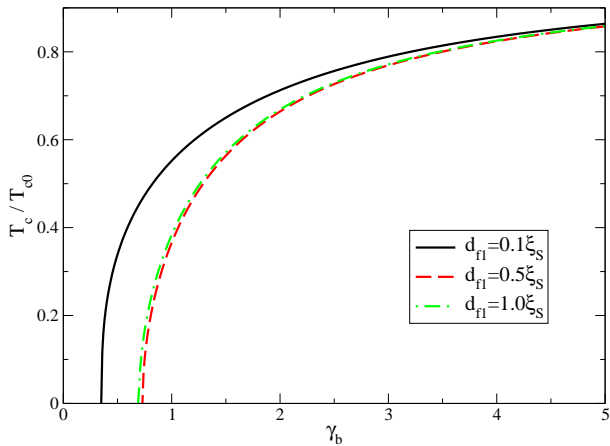}
\caption{Critical temperature $T_c$ of a trilayer versus the interface
resistance parameter $\gamma_b=\gamma_{b1}=\gamma_{b2}$ at a few
thicknesses $d_{f1}$ corresponding to points on the $J=20T_{c0}$ curve
in Fig.~\ref{Fig_tri_Tc-vs-df1-and-J} to the left, inside, and to the
right of the $T_{c}=0$ region. The other parameters are the same as in
Fig.~\ref{Fig_tri_Tc-vs-df1-and-J}.}
\label{Fig_tri_Tc-vs-gammaB-and-df1}
\end{figure}

\begin{figure}[th]
\includegraphics[width=8cm]{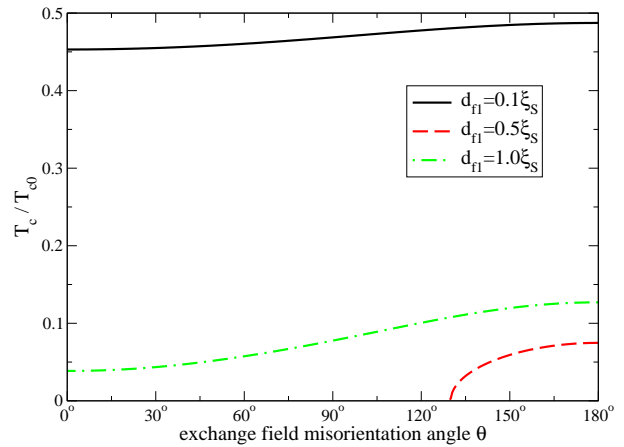}
\caption{Critical temperature $T_c$ of a trilayer versus the
misorientation angle $\theta$ between the exchange fields in
ferromagnets F$_1$ and F$_2$. The exchange field is $J_1=J_2=20T_{c0}$
and the other parameters are the same as in
Fig.~\ref{Fig_tri_Tc-vs-df1-and-J}.}
\label{Fig_tri_Tc-vs-theta-and-df1}
\end{figure}

\begin{figure}[th]
\includegraphics[width=8cm]{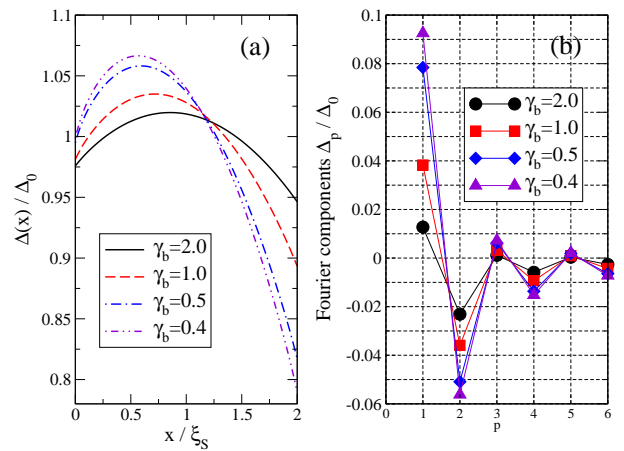}
\caption{(a) The spatial dependence of the order parameter for several
interface resistance parameters $\gamma_b=\gamma_{b1}=\gamma_{b2}$ on
the solid curve ($d_{f1}=0.1\xi_S$) in
Fig.~\ref{Fig_tri_Tc-vs-gammaB-and-df1}. (b) The Fourier components in
Eq.~(\ref{fourier}). Note that $\Delta$ is normalized to the first
component $\Delta_0$, which remains unknown in a linearized theory.}
\label{Fig_tri_Delta-vs-gammaB}
\end{figure}

The exact point where superconductivity disappears depends on many
other parameters in addition to the strength of the exchange
field. For example, the influence of the second ferromagnet's
thickness $d_{f2}$ can be understood in terms of the initial $T_c$
suppression at $d_{f1}=0$, which corresponds to the bilayer case. For
sufficiently large $d_{f2}$ the initial suppression is large enough
that the subsequent oscillations with increasing $d_{f1}$ lead to
disappearing and reappearing superconductivity, see
Fig.~\ref{Fig_tri_Tc-vs-df1-and-df2}. Naturally, the initial $T_c$
suppression at $d_{f1}=0$ is a non-monotonic function of $d_{f2}$, in
analogy to the $T_{c}(d_{f1})$-dependence.

Another important parameter for the size of the $T_c$ variations, is
the interface resistance. As seen in
Fig.~\ref{Fig_tri_Tc-vs-gammaB-and-df1}, superconductivity is
suppressed in trilayers with good contacts (small $\gamma_b$) for the
model parameters chosen here. For example it is not possible to
consider $d_s\ll\xi_s$, for which simplified calculations with a
constant $\Delta$ can be made, and simultaneously consider good
contacts $\gamma_b\rightarrow 0$ for reasonable conductivity
mismatches (here $\gamma=0.3$). Certainly, for larger conductivity
mismatches (smaller $\gamma$), $T_c$ is not suppressed as much and a
smaller $\gamma_b$ can be used. However, it is always important to
keep in mind that $T_c$ is suppressed to zero in quite a large
parameter space, including small $d_s$ and small $\gamma_b$.

In Fig.~\ref{Fig_tri_Tc-vs-theta-and-df1} we show the influence of the
relative direction of the exchange fields in the two ferromagnetic
layers. The dependence is monotonic, with the parallel orientation
being the most destructive. We note (see also
Ref.~\onlinecite{Fom2003}) that for parallel or antiparallel exchange
field orientations triplet correlations with zero spin projection on
the local exchange field are present in the structure, while for
intermediate orientations triplet correlations with non-zero
projection are also induced. In order to describe the
$\theta$-dependence correctly, it is therefore important to include
${\vec f}_t$, see Appendix~\ref{App_trilayer}.

In Fig.~\ref{Fig_tri_Delta-vs-gammaB}(a) we present order parameter
profiles for four different values of $\gamma_b$ on the solid line
($d_{f1}=0.1\xi_S$) in Fig.~\ref{Fig_tri_Tc-vs-gammaB-and-df1}. For a
good contact (small resistance $\gamma_b$), the pair breaking becomes
quite severe. The suppression is reflected as a growth of the Fourier
components $p \geq 1$, see Fig.~\ref{Fig_tri_Delta-vs-gammaB}(b).

\section{Pentalayer}\label{Sec_ResultsPentalayer}

\begin{figure}[t]
\includegraphics{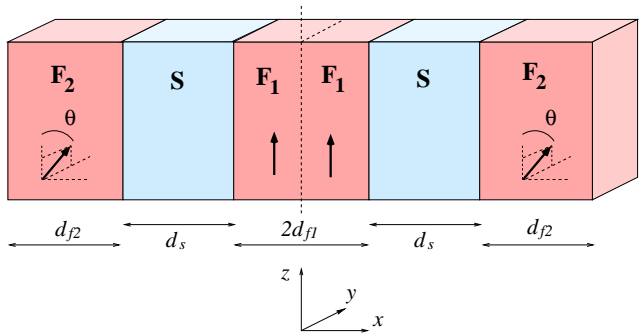}
\caption{The F$_{2}$/S/F$_{1}$/S/F$_{2}$ pentalayer structure. We
consider two types of misalignment of the outer exchange fields
relative to the exchange field in the center layer. First, as shown
here, $J_2$ is rotated by the same angle $\theta$. The second
possibility is when $J_2$ is rotated in opposite directions, $-\theta$
in the left F$_2$ and $+\theta$ in the right F$_2$.}
\label{Fig_penta}
\end{figure}

Consider the pentalayer shown in Fig.~\ref{Fig_penta}. Experimental
results for the critical temperature, including signatures of a
transition from a zero- to a $\pi$-junction as function of the
thickness of the central F-layer, were recently presented for this
structure in Ref.~\onlinecite{Shelukin}. The theory fits of $T_c$ of
the pentalayers in Ref.~\onlinecite{Shelukin} were obtained with the
theory presented in the present paper.

The superconducting layers are considered geometrically identical with
identical bulk material critical temperatures $T_{c0}$. In the central
ferromagnetic layer (F$_{1}$), the exchange field is aligned with the
$z$-axis, while in the right and left layers (F$_{2}$) it forms an
angle $\theta$ with respect to the $z$-axis. We characterize the
different layers by their thicknesses, exchange fields, and diffusion
constants, with the constraint that the pentalayer should have certain
symmetries with respect to the midpoint, see below. The present
pentalayer problem can then be reduced to a trilayer problem with a
new effective boundary condition at a fictitious outer surface at the
center ($x=0$). The two superconducting order parameters in the left
and right S layers may differ in phase, which is reflected in the
effective boundary condition.

We shall consider two types of misorientation of the exchange fields
in the outer layers relative to the center layer: the exchange fields
$J_2$ are rotated by $+\theta$ as in Fig.~\ref{Fig_penta} (rotation
type 1, $+\theta/+\theta$), or rotations by $-\theta$ and $+\theta$ in
the left and right outer layers respectively (rotation type 2,
$-\theta/+\theta$).

For rotation type 1 ($+\theta/+\theta$), when the phase difference
vanishes (0-junction case), the singlet component $f_{s}$ is an even
function of $x$.  Considering the parity of the exchange field ${\bf
J}$ ($J_{z} \to J_{z}$ and $J_{y} \to J_{y}$) and the Eqs.
(\ref{coup1})-(\ref{coup2}), we deduce that the $f_{tz}$ and $f_{ty}$
components have the same even parity. Thus, we impose the conditions
\begin{equation}
(+\theta,+\theta),\; 0-\mathrm{jct}: f'_{s}(0)=f'_{tz}(0)=f'_{ty}(0)=0. \label{bound10}
\end{equation}
On the other hand, when the phase difference is $\pi$ ($\pi$-junction
case), $f_{s}$, $f_{tz}$ and $f_{ty}$ are odd functions of $x$ and we
impose the conditions
\begin{eqnarray}
(+\theta,+\theta),\; \pi-\mathrm{jct}: f_{s}(0)=f_{tz}(0)=f_{ty}(0)=0.
\label{bound1pi}
\end{eqnarray}

For rotation type 2 ($-\theta/+\theta$), the exchange field component
$J_{y}$ is instead odd under $x\to -x$. For the 0-junction case, it
implies that $f_{ty}$ is odd, while the other components are even just
as above. For the $\pi$-junction case the parities are
interchanged. The effective boundary conditions are
\begin{eqnarray}
(-\theta/+\theta),\; 0-\mathrm{jct}: && \left\{
\begin{array}{l}
f'_{s}(0)=f'_{tz}(0)=0,\\
f_{ty}(0)=0,
\end{array}
\right.\label{bound20}\\
(-\theta/+\theta),\; \pi-\mathrm{jct}: && \left\{
\begin{array}{l}
f_{s}(0)=f_{tz}(0)=0,\\
f'_{ty}(0)=0.
\end{array}
\right.\label{bound2pi}
\end{eqnarray}

As shown in Appendix~\ref{App_pentalayer}, the different boundary
conditions yield different matrices $\hat{W}$ for the effective
boundary condition (\ref{bound}).

\begin{figure}[t]
\includegraphics[width=8cm]{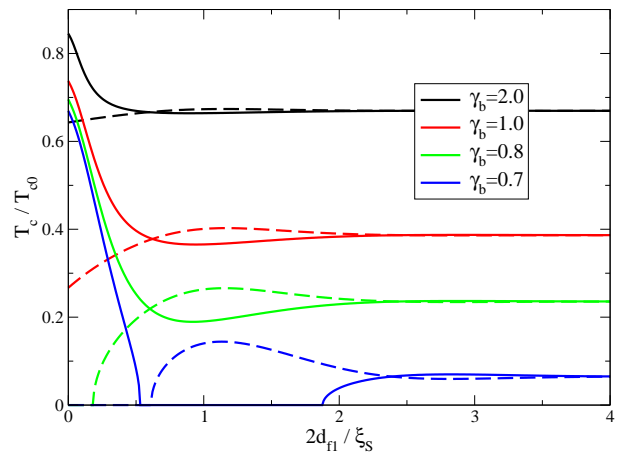}
\caption{Critical temperature $T_c$ of a pentalayer versus the center
ferromagnet layer thickness $2d_{f1}$ for several barrier
transparencies. The curves come in pairs, solid line for the
0-junction and dashed line for the $\pi$-junction, from top to bottom
for $\gamma_{b1}=\gamma_{b2}=\gamma_b=\{2,1,0.8,0.7\}$
respectively. The other parameters are $d_S=2\xi_S$,
$d_{f2}=0.5\xi_S$, $J_1=J_2=20T_{c0}$, $\theta=0$,
$\gamma_1=\gamma_2=0.3$, and $D_{f1}=D_{f2}=D_S$.}
\label{Fig_penta_Tc-vs-df1-and-gammaB}
\end{figure}

\begin{figure}[th]
\includegraphics[width=8cm]{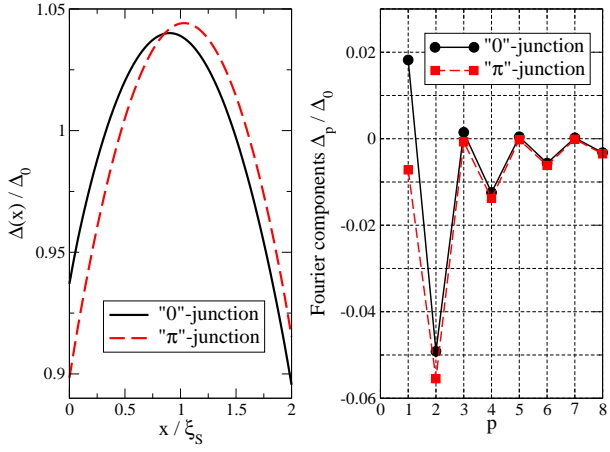}
\caption{(a) The order parameter profile for the parameters in
Fig.~\ref{Fig_penta_Tc-vs-df1-and-gammaB} for $\gamma_b=0.8$ at
$2d_{f1}=0.4\xi_S$, for which the critical temperature for the 0- and
$\pi$-junctions are respectively $T_{c}\approx 0.3T_{c0}$ and
$T_c=0.16T_{c0}$. (b) The corresponding Fourier components.}
\label{Fig_penta_Delta}
\end{figure}

\begin{figure}[th]
\includegraphics[width=8cm]{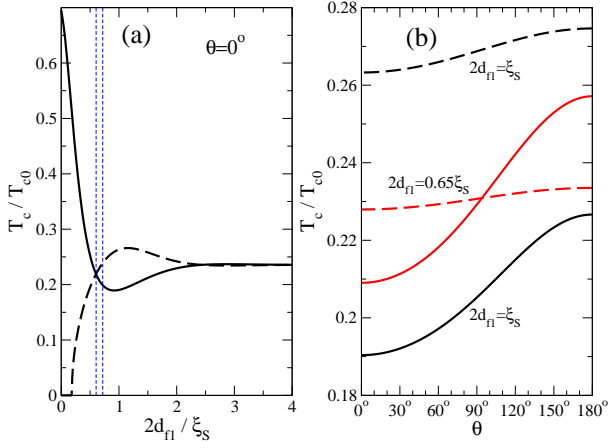}
\caption{(a) The same curves as in
Fig.~\ref{Fig_penta_Tc-vs-df1-and-gammaB} for $\gamma_b=0.8$. For
thicknesses $d_{f1}$ between the two vertical lines the $0\to\pi$
transition can be tuned by the relative orientation of the exchange
fields in F$_1$ and F$_2$. (b) The switch $0\to\pi$ appears at a
critical angle $\theta_c\approx 84^o$ for the thickness
$2d_{f1}=0.65\xi_S$. For the larger thickness $2d_{f1}=\xi_S$, outside
the window indicated in (a), the largest $T_c$ is obtained for the
$\pi$-junction.}
\label{Fig_penta_Tc-vs-df1-and-theta}
\end{figure}

\subsection{Results}

The dependence of $T_c$ on the various parameters in the model is
similar for the trilayer with left ferromagnetic layer thickness
$d_{f1}$ and for the pentalayer with a phase difference $0$ between
the two superconductors and a central ferromagnet layer thickness
$2d_{f1}$. The critical temperature is in fact equal for rotation type
I, since the boundary condition at the center of the pentalayer,
Eq.~(\ref{bound10}), is the same as for the outer surface of the
trilayer. Note, however, that the boundary condition for one of the
triplets is different for rotation type II, see
Eq.~(\ref{bound20}). The new ingredient in the pentalayer case is the
possibility of a phase difference $\pi$ between the
superconductors. The $\pi$-state can in simplified terms be understood
as being due to the oscillatory behavior of the Green's function
$f_s(\varepsilon_n,x)$ inside the central ferromagnetic layer
F$_1$. In an experiment, $T_c$ is given by the largest $T_c$ for each
thickness and there will be a sudden almost kink-like change in $T_c$
at the $0\to\pi$ transition. For large oscillations, the transition
becomes sharper. This is illustrated by changing the interface
resistance $\gamma_b$ in
Fig.~\ref{Fig_penta_Tc-vs-df1-and-gammaB}. For good contacts and
strong exchange fields, superconductivity can be destroyed at some
critical thickness and then reappear at a larger thickness, just as in
the trilayer case. For the pentalayer, however, the $\pi$-phase can
pre-empt the $0$-phase and superconductivity appears earlier compared
to the trilayer as $d_{f1}$ is increased, see the curves for
$\gamma_b=0.7$ in Fig.~\ref{Fig_penta_Tc-vs-df1-and-gammaB}.

\begin{figure}[th]
\includegraphics[width=8cm]{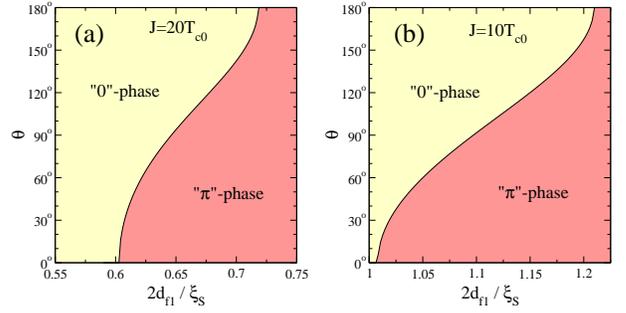}
\caption{(a) Phase diagram of the $0\to\pi$ transition in the window
  indicated in Fig.~\ref{Fig_penta_Tc-vs-df1-and-theta}(a). In (b) we
  show the phase diagram for a smaller exchange field, $J=10T_{c0}$.
  The range of thicknesses $d_{f1}$ for which there is a switching by
  changing $\theta$ is larger in this case.}
\label{Fig_penta_thetaC-vs-df1}
\end{figure}

\begin{figure}[th]
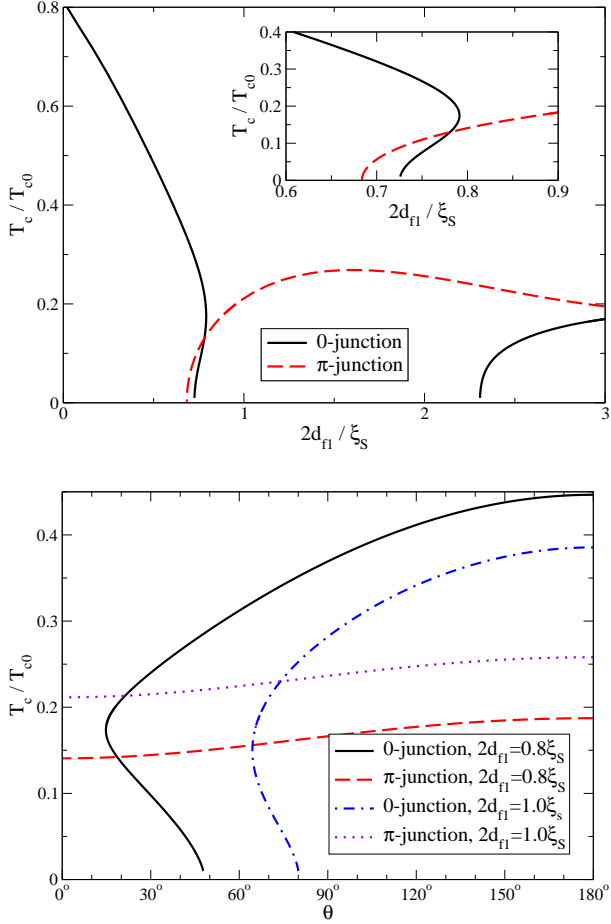

\includegraphics[width=8cm]{Fig13a_penta_backbend.eps}\vspace{0.5cm}
\includegraphics[width=8cm]{Fig13b_penta_backbend2.eps}\vspace{0.5cm}
\caption{Upper panel: In the region where $T_c$ is suppressed to zero,
the curve $T_c(d_{f1})$ can contain a back-bend. This latter could
signal the occurrence of a first-order transition, which is however
beyond the scope of the present theory. For the pentalayer, the
$\pi$-phase can interfere and the first order transition might be
avoided. The parameters are $d_s=2\xi_s$, $d_{f2}=0.2\xi_s$,
$\gamma_1=\gamma_2=0.35$, $\gamma_{b1}=\gamma_{b2}=0.4$,
$J_1=J_2=10T_{c0}$ and $\theta=0$. In the lower panel we study the
dependence on the exchange field for two particular thicknesses
$d_{f1}$ in the upper panel. Clearly, the back-bend behavior can occur
also as function of the exchange field misorientation angle $\theta$.}
\label{Fig_penta_backbend}
\end{figure}

\begin{figure}[th]
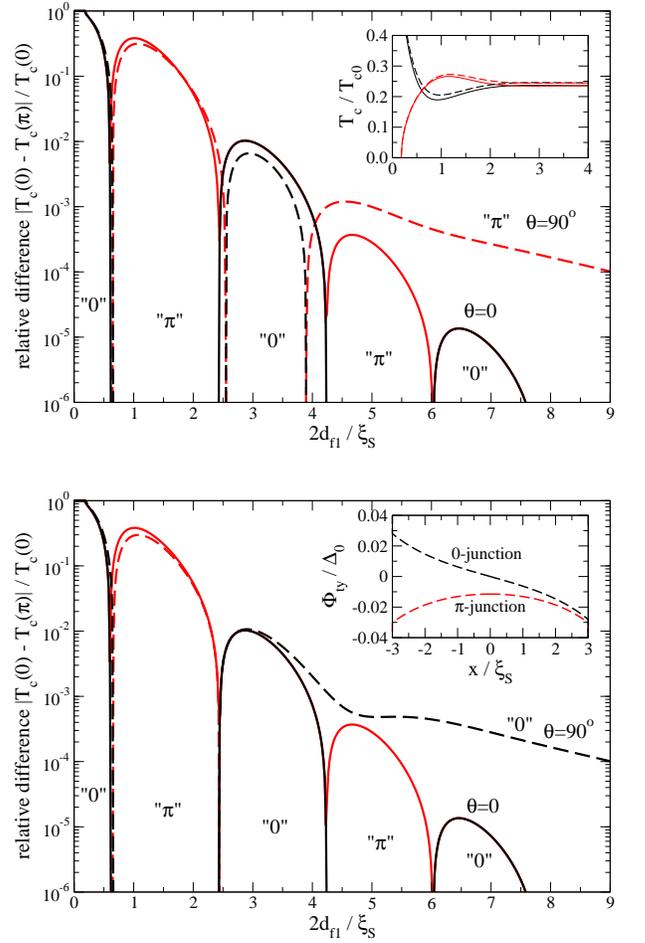

\includegraphics[width=8cm]{Fig14a_penta_LongRangeTcDifference.eps}\vspace{0.5cm}
\includegraphics[width=8cm]{Fig14b_penta_LongRangeTcDifferenceB.eps}
\caption{For thick center films (large $d_{f1}$) the communication
between the two superconductors is taken over by long-range
non-oscillatory equal spin triplet correlations and the junction is
stabilized at phase difference 0 or $\pi$ depending on the exchange
field orientation: the upper panel ($\pi$-junction for $\theta=90^o$)
is obtained for exchange field rotation type 1 ($+\theta$ in both the
left and right outer ferromagnets F$_2$, as illustrated in
Fig.~\ref{Fig_penta}), while the lower panel (0-junction for
$\theta=90^o$) is obtained for rotation type 2 ($-\theta$ in the left
F$_2$ and $+\theta$ in the right F$_2$). The difference in $T_c$
between the $0$ and $\pi$ cases is quite small for large
$d_{f1}$. Upper inset: the differences in $T_c$ for various exchange
field orientations (solid lines $\theta=0$, dashed lines
$\theta=90^o$) are due to the interaction between the ferromagnetic
layers $F_{1}$ and $F_2$. Lower inset: spatial dependence inside the
central $6\xi_s$ thick F$_1$ layer of the long-range triplet $f_{ty}$
induced for $\theta=90^o$. The parameters are $d_s=2\xi_s$,
$d_{f2}=0.5\xi_s$, $d_{f1}=3\xi_s$, $\gamma_1=\gamma_2=0.3$,
$\gamma_{b1}=\gamma_{b2}=0.8$ and $J_1=J_2=10T_{c0}$.}
\label{Fig_penta_LongRangeTcDifference}
\end{figure}

An example of the order parameter suppression is shown in
Fig.~\ref{Fig_penta_Delta}(a), corresponding to $\gamma_b=0.8$ and
$2d_{f1}=0.4\xi_S$ in Fig.~\ref{Fig_penta_Tc-vs-df1-and-gammaB}. The
suppression of $\Delta$ at the interfaces is more severe for phase
difference $\pi$ and the 0-junction is stabilized, i.e. has the
largest $T_c$ as seen in Fig.~\ref{Fig_penta_Tc-vs-df1-and-gammaB}.

In the region close to the $0\to\pi$ transition, it is possible to
switch between the $0$- and $\pi$-phases by changing the relative
orientation of the exchange fields. We note that this possibility was
already deduced from calculations of the Josephson critical current in
the papers \onlinecite{Gol2002} and \onlinecite{Ber2003} considering
different geometries.  We illustrate this effect in
Fig.~\ref{Fig_penta_Tc-vs-df1-and-theta}: switching is possible in
between the vertical lines in
Fig.~\ref{Fig_penta_Tc-vs-df1-and-theta}(a). Since experimentally,
$T_c$ is given by the largest $T_c$ for each $\theta$, the $0\to\pi$
switch would show up as a sudden almost kink-like change in $T_c$ with
the variation of $\theta$, as shown in
Fig.~\ref{Fig_penta_Tc-vs-df1-and-theta}(b).  We present in
Fig.~\ref{Fig_penta_thetaC-vs-df1} the phase diagram of the junction
in the region around the window indicated in
Fig.~\ref{Fig_penta_Tc-vs-df1-and-theta}(a). The window inside which a
$0\to\pi$ phase change can be induced by the orientation angle
$\theta$ is larger for a smaller exchange field since the
$T_c$-oscillation period is longer in this case. We see this effect by
comparing the $J=20T_{c0}$ case in
Fig.~\ref{Fig_penta_Tc-vs-df1-and-theta}(a) to the $J=10T_{c0}$ case
shown in (b).

It has been found\cite{Fom2002,Bal2003,Buz} for the bilayer and
trilayer cases that $T_c$ can become a multiple valued function of
e.g. the thickness of the ferromagnet. We show this type of behavior
for the pentalayer case in Fig.~\ref{Fig_penta_backbend} (upper
panel). The non-monotonic dependence of $T_c$ is similar to the case
of a clean thin film in an in-plane magnetic field\cite{FFLO1,FFLO2}
and to thin films of superfluid $^3\mathrm{He}$
[\onlinecite{He-3-FFLO1,He-3-FFLO2}]. For these clean systems it has
been proposed that an inhomogeneous superconducting state can be
formed. In a dirty system, such inhomogeneity seems very unlikely and
it has instead been proposed that the back-bend signals the
possibility of a first-order transition in the system.\cite{Buz}
First-order transitions are however beyond the scope of the present
paper. Instead we point out that for the pentalayer case, the
$\pi$-phase becomes favorable in the same region of thicknesses as
where there is a back-bend for the 0-junction. The back-bend behavior
for the $0$-junction, and the interfering $\pi$-phase, occurs also as
function of the exchange field misorientation angle $\theta$, see the
lower panel in Fig.~\ref{Fig_penta_backbend}. Interestingly, there is
a discontinuous drop in $T_c$ at the $0\to\pi$ transition when
$\theta$ is tuned from around $20^o$ down to $10^o$, see the solid and
dashed lines in Fig.~\ref{Fig_penta_backbend}.

For very large thicknesses $d_{f1}$ the predominant superconducting
correlations that penetrate F$_1$ and connect the two superconductors
are the long-range non-oscillatory triplet components of ${\bf
f}_t$. As a consequence, at large $d_{f1}$, $T_{c}$ becomes a
monotonic function of $d_{f1}$. The difference in $T_c$ between the
$0$- and $\pi$-phases is, however, quite small, see
Fig.~\ref{Fig_penta_LongRangeTcDifference}. The junction is stabilized
at large $d_{f1}$ either at 0 or at $\pi$ phase difference, depending
on the way the exchange fields of the two outer ferromagnetic layers
are rotated relative to the center layer (a similar effect associated
with the chirality of the rotation has been found in
Ref.~\onlinecite{Ber2003} from calculations of the critical current in
S-F multilayered junctions). We consider two types of rotation: the
exchange fields in the outer ferromagnets are rotated by $+\theta$ as
in Fig.~\ref{Fig_penta} (rotation type 1, $+\theta/+\theta$), or
rotations by $-\theta$ and $+\theta$ in the left and right outer
layers respectively (rotation type 2, $-\theta/+\theta$). The only
difference between the two phase differences 0 and $\pi$ is the parity
property of one of the triplets: $f_{ty}$ is an even (odd) function of
$x$ for the 0-junction ($\pi$-junction) for rotation type 1
($+\theta/+\theta$). The parity properties of $f_{ty}$ for $0$ and
$\pi$ phase differences are reversed for rotation type 2
($-\theta/+\theta$). When $f_{ty}$ has odd parity it is smaller
compared to the even parity case, which leads to a smaller suppression
of the singlet $f_s$, i.e. less pair breaking, and a higher $T_c$. We
therefore have a $\pi$-junction at large $d_{f1}$ for rotation type 1
(upper panel in Fig.~\ref{Fig_penta_LongRangeTcDifference}), and a
$0$-junction for rotation type 2 (lower panel in
Fig.~\ref{Fig_penta_LongRangeTcDifference}). We show the spatial
dependence of the long range (in the central $F_1$ layer) triplet
Green's function
\begin{equation}
\Phi_{ty}(x) = T\sum_{\varepsilon_n>0} f_{ty}(\varepsilon_n,x)
\end{equation}
in the lower inset of Fig.~\ref{Fig_penta_LongRangeTcDifference}

Experimentally, the transition from $0 \to \pi$ was studied until now
by varying the thickness
\cite{Kon2002,Blu2002,Gui2003,Obi2005,Shelukin} of the ferromagnet in
S-F-S junctions, or by varying the
temperature,\cite{Rya2001,Sel2004,Fro2004} which is more practical
since the transition is seen in the same device. Here we have studied
another possibility to switch from the 0 to the $\pi$ state within the
same device, namely by continuously changing the relative orientation
of the ferromagnetic moments. Our results are qualitatively consistent
with the results obtained within Josephson critical current
calculations.\cite{Gol2002,Ber2003} The feasibility of controlling the
orientation of the moments has been proven experimentally through the
investigation of F-S-F trilayers for different moment
orientations.\cite{Gu2002,Pot2005,Mor2006,Rus2006}

\section{Discussion of the Numerics}\label{Sec_Numerics}

Let us discuss some delicate problems that need to be addressed when
$T_c$ is computed in inhomogeneous structures. In particular, we will
compare the two methods of computing $T_c$: Eq.~(\ref{eigenp}) which
we call the Fourier method and Eq.~(\ref{self}) which we call the grid
method.

\begin{figure}[t]
\includegraphics[width=8cm]{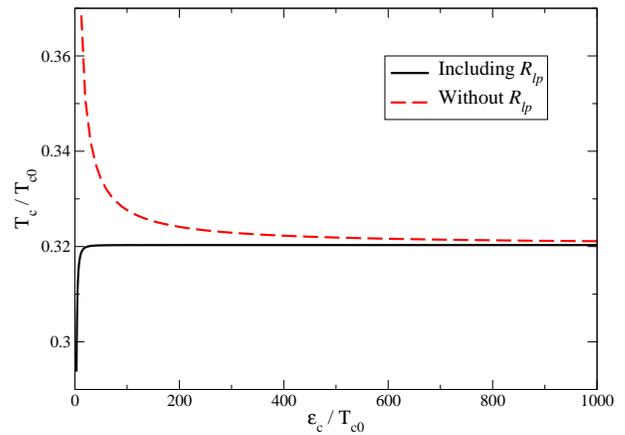}\vspace{0.5cm}
\includegraphics[width=8cm]{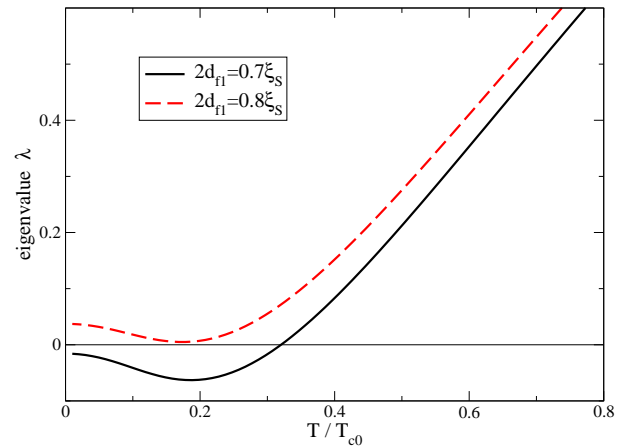}
\caption{Upper panel: critical temperature versus the technical
cut-off $\varepsilon_c$. To achieve good accuracy for the critical
temperature we need $\varepsilon_c$ of order $1000T_{c0}$ (dashed
line). When the high-energy tail is summed to infinity, as described
for the Fourier method in Eq.~(\ref{Rlp}), the convergence is more
acceptable (solid line). Lower panel: the eigenvalue of the gap
equation (\ref{eigenp}) versus temperature for two different
thicknesses. The zero-crossing determines $T_c$. When $T_c$ is
suppressed, $\lambda(T)$ can become a flat function of $T$ which makes
it important to compute $\lambda$ with high accuracy to avoid
numerical errors in $T_c$. The parameters in the upper panel were
chosen as in Fig.~\ref{Fig_penta_backbend}, 0-junction, at
$2d_{f1}=0.7$ and $\theta=0$.  In the lower panel the two thicknesses
are indicated in the legend.}
\label{Fig_numerics}
\end{figure}

The most important problem to address in any calculation using
Usadel's approximation is the fact that the Matsubara sum in
Eq.~(\ref{gap}) is intrinsically slowly convergent, as compared to
calculations done with the more general Eilenberger approach. As we
show in Appendix~\ref{App_HE}, the difference
$f_s(\varepsilon_n)-\pi\Delta/|\varepsilon_n|$ appearing in the gap
equation (\ref{gap}) is at high-energies proportional to
$1/\varepsilon_n^2$ for inhomogeneous systems. This can be contrasted
with an Eilenberger approach, where the high-energy asymptotic is
$1/|\varepsilon_n|^3$. It is therefore always necessary to extend the
Matsubara sum to high energies when the Usadel approximation is
employed, see the dashed line in Fig.~\ref{Fig_numerics}. In the
example we need a technical cut-off of order $1000T_{c0}$ to compute
$T_c$ with an accuracy of $1\%$. However, since the high-energy form
of $f_s(\varepsilon_n)$ is known (see Appendix~\ref{App_HE}) it is in
principle possible to circumvent the problem by treating the
high-energy tail separately and sum the Matsubara sum to infinity. We
have done that within the Fourier series approach, see
Appendix~\ref{App_HEfourier}. A more acceptable cut-off of order
$100T_c$ is then enough to achieve excellent accuracy, see the solid
line in Fig.~\ref{Fig_numerics}.

There are several other factors that, together with the slow
convergence of the Matsubara sum, conspire to make it non-trivial to
achieve acceptable accuracy, especially when $T_{c}$ is small compared
to $T_{c0}$. The critical temperature is computed by finding the
temperature for which the eigenvalue $\lambda$ of the gap equation is
zero [Fourier method, Eq.~(\ref{eigenp})] or one [grid method,
Eq.~(\ref{self})]. The function $\lambda(T)$ can become a very flat
function of $T$ in the region where $T_{c0}$ is small, see the lower
panel of Fig.~\ref{Fig_numerics}. Any error made in the calculation of
$\lambda$ can therefore be magnified to a larger error in $T_c$ and it
becomes increasingly critical to compute $\lambda$ with high accuracy
as $T_{c}$ is suppressed.

The above two technical problems are particularly hard to circumvent
within the grid method. First of all, the need to include high
energies up to a technical cut-off $\varepsilon_c$ imposes a condition
on the grid spacing $\delta x$. At high energies the function
$G(\varepsilon_n,x,y)$ is typically peaked in the region $x\sim y$
\begin{equation}
G(\varepsilon_n\gg T_{c0},x,y) \sim \frac{k_s(\varepsilon_n)}{\varepsilon_n} e^{-k_s(\varepsilon_n)|x-y|},
\label{G_HE}
\end{equation}
where $k_s(\varepsilon_n)=\sqrt{2\varepsilon_n/D}$. It is therefore
necessary to choose
\begin{equation}
\frac{\delta x}{\xi_S} \ll \frac{1}{\xi_S k_s(\varepsilon_c)} =\sqrt{\frac{\pi T_{c0}}{\varepsilon_c}},
\end{equation}
to resolve this dependence. Since we need a cut-off around
$1000T_{c0}$, because of the slow convergence within the Usadel
approach, we need a grid spacing of order $0.01\xi_S$ or finer. The
matrix in Eq.~(\ref{self}) must therefore typically be of the order of
a few hundred elements square, which severely slows down the numerics.

One reason for the importance to resolve the peaked form of
$G(\varepsilon_n,x,y)$ is due to the interchange in order of the
Matsubara sum and the integration over $y$ in Eq.~(\ref{self}). We
write Eq.~(\ref{self}) as
\begin{equation}
\int_{0}^{d_{s}} K(x,y)\Delta(y) dy= \Delta(x),
\end{equation}
and compute each element of the matrix $K(x,y)$ by summing over
$\varepsilon_n$. The asymptotic form of the diagonal is however
$G(\varepsilon_n,x,x)\propto 1/\sqrt{\varepsilon_n}$ and the Matsubara
sum is not convergent. This is in principle irrelevant for the
calculation of $T_c$ because $T_c$ only depends on the eigenvalue of
the matrix, which is a quantity given by the Matsubara sum integrated
over $y$. Note that when Eq.~(\ref{G_HE}) is integrated over $y$, a
factor $1/k_s$ appears in the primitive function of the exponential
and the asymptotic form is $1/|\varepsilon_n|$, which is (by
construction) cancelled by the sum over $1/|\varepsilon_n|$ in the
denominator of $K(x,y)$, see Eq.~(\ref{self}). Numerically, however,
the integral over the discretized coordinate $y$ can only be computed
with some accuracy given by the grid spacing $\delta x$. The error
made in computing the integral is transferred into an error in the
eigenvalue $\lambda$ which, as described above, can result in an error
in $T_c$ magnified by the flatness of the $\lambda(T)$-dependence.

To circumvent the problems described above, one must predict the
high-energy tail to avoid cut-offs larger than $\sim
100T_{c0}$. Within the grid-method that means computing the derivative
of $\Delta(x)$, i.e. to introduce an approximate formula for the
derivative on a discretized grid. But that also introduces numerical
errors and the grid must still be dense, which means that the matrix
$K(x,y)$ remains large and the calculation with the grid method is
always very slow and susceptible to numerical errors.

\begin{figure}
\includegraphics[width=8cm]{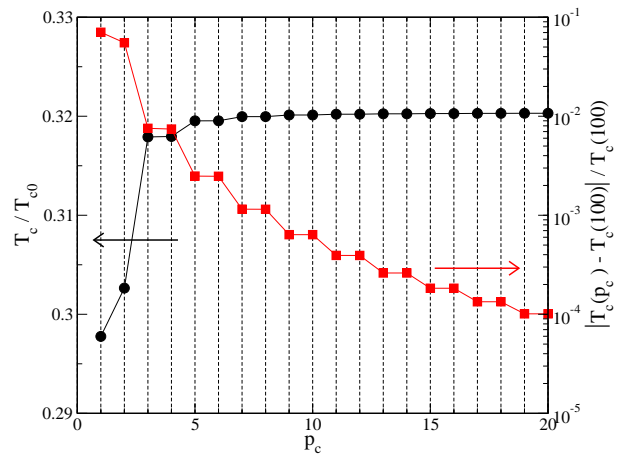}
\caption{Critical temperature versus the number $p_c$ of included
Fourier coefficients in Eq.~(\ref{fourier}). The variation in absolute
numbers are shown by circles (vertical scale to the left), while the
variation in relation to the corresponding value for $T_c$ at a high
cut-off $p_c=100$ is shown by squares (vertical scale to the
right). The even-odd variation is due to the choice of parameters: the
junction is almost symmetric and the even-number Fourier components
corresponding to symmetric $\cos$-functions contribute more to
$T_c$. The model parameters were chosen as in Fig.~\ref{Fig_numerics},
upper panel.}
\label{Fig_numerics_pc}
\end{figure}

All the problems related to the discretization of the spatial
coordinate are avoided within the Fourier series approach, since
$G(\varepsilon_n,x,y)$ is analytically integrated over $x$ and $y$ in
the course of the derivation of the matrix $m_{lp}$ in
Eq.~(\ref{eigenp}), see Appendix~\ref{App_fourier}. Moreover, the
high-energy tail is easily predicted analytically, see
Appendix~\ref{App_HEfourier}. It is typically sufficient to include
only the 20 first Fourier components in the calculation of $T_c$, see
Fig.~\ref{Fig_numerics_pc}. The matrix $m_{lp}$ is therefore small,
the high-energy cut-off of the Matsubara sum can be chosen reasonably
small, and very high accuracy is achieved while the speed of the
calculation remains very high.

\section{Summary}\label{Sec_Summary}

In conclusion, we have studied the change of the superconducting critical
temperature, $T_{c}$, in asymmetric 
trilayers F$_{1}$-S-F$_{2}$ and symmetric pentalayers
F$_{2}$-S-F$_{1}$-S-F$_{2}$ with any relative orientation angle
between the magnetizations of F$_{1}$ and F$_{2}$. 
For both cases we have presented phase diagrams, showing $T_c$ as function
of the misorientation angle, $\theta$,  and as a function of the 
ferromagnet layer thicknesses.
We have investigated the interplay of long-range triplet components and
Josephson coupling in the pentalayer geometry.  
We have demonstrated the possibility to switch between
the 0 and $\pi$ states by controlling the relative orientation of the
F moments in a pentalayer structure. This behavior may be appealing
for the experimental study of the $0 \to \pi$ transition.
We have presented details for a general method for the computation
of $T_{c}$ and the dependence
of the order parameter on the spatial coordinates in diffusive hybrid
structures.  With this technique, the accuracy as well as the speed of
the numerics are immensely improved compared with previously used
techniques.

\acknowledgments

We acknowledge support from the Deutsche Forschungsgemeinschaft within
the Center for Functional Nanostructures (T.C., M.E.), and the
Alexander von Humboldt Foundation (T.L.).

\appendix

\section{Derivation of $\hat W$ for the Trilayer}\label{App_trilayer}

In this Appendix we provide the details of the calculations leading to
the effective boundary condition (\ref{bound}) obeyed by the singlet
component in the superconducting region in asymmetric
F$_{1}$-S-F$_{2}$ trilayers with an arbitrary mutual orientation
between the magnetizations in F$_{1}$ and F$_{2}$. Except for this
latter component, it is possible to derive analytically the spatial
dependences of all the components of the anomalous Green's function
$f$ close to $T_{c}$ (next Section). In Section \ref{SecW} we
determine from the consideration of the boundary conditions
(\ref{bound1})-(\ref{bound2}) the matrix $\hat{W}$ that enters the
expression for the effective boundary condition (\ref{bound}).

\subsection{Spatial Dependences}

In the superconducting layer, the triplet vector ${\bf f}_{t}$ obeys a
homogeneous differential equation [Eq. (\ref{coup2}) with ${\bf J}
={\bf 0}$] which is straightforwardly solved:
\begin{eqnarray}
{\bf f}_{t}= {\bf c} \cosh(k_{s}x)+ {\bf d} \sinh(k_{s}x) \label{expS}
\end{eqnarray}
with ${\bf c}$ and ${\bf d}$ constants.

For a fixed exchange field in each F layer, the system of coupled
Eqs. (\ref{coup1})-(\ref{coup2}) can be easily solved in the
ferromagnetic regions. After application of the boundary conditions at
the outer surfaces, the solutions can be written in the
form\cite{Cha2005}
\begin{widetext}
\begin{eqnarray}
\left( \begin{array}{c} f_{s} \\{\bf f}_{t}
\end{array}
\right)
= \sum_{\varepsilon=\pm} a_{\varepsilon} \cosh\left[k_{\varepsilon \, 1} (x+d_{f1}\right]
\left(\begin{array}{c} 1\\
\varepsilon \hat{{\bf z}}
\end{array}
\right) 
+ a_{0} \cosh\left[k_{0 \, 1} (x+d_{f1}) \right] \left(\begin{array}{c} 0\\
\hat{{\bf y}}
\end{array}
\right) \label{expF10j}
\end{eqnarray}
for the F$_{1}$ layer, and
\begin{eqnarray}
\left( \begin{array}{c} f_{s} \\{\bf f}_{t}
\end{array}
\right)
= \sum_{\varepsilon=\pm}
 b_{\varepsilon} \cosh\left[k_{\varepsilon \, 2} (x-d_{s}-d_{f2})\right]
\left(\begin{array}{c} 1\\
\varepsilon (\cos \theta \,\hat{{\bf z}}+ \sin \theta \,\hat{{\bf y}})
\end{array}
\right) 
+ b_{0} \cosh\left[k_{0 \, 2} (x-d_{s}-d_{f2}) \right] \left(\begin{array}{c} 0\\
\sin \theta \,\hat{{\bf z}}- \cos \theta \,\hat{{\bf y}}
\end{array}
\right) \label{expF2}
\end{eqnarray}
for the F$_{2}$ layer. Here we have defined
\begin{eqnarray}
k_{\pm \, q}&=&\sqrt{(2\varepsilon_{n} \pm 2i J_{q})/D_{fq}}, \\
k_{0 \, q}&=&\sqrt{2 \varepsilon_{n}/D_{fq}},
\end{eqnarray}
with the index $q=1$ or 2 referring to the F$_{1}$ or F$_{2}$ layer.

\subsection{Determination of $\hat{W}$} \label{SecW}

The constants $a_{j}$ and $b_{j}$ ($j=\pm,0$), ${\bf c}$ and ${\bf d}$
are determined with the help of the boundary conditions
(\ref{bound1})-(\ref{bound2}) considered for the two S/F interfaces.
Writing these conditions for the triplet components only, we have
\begin{eqnarray}
\xi_{s} f'_{l} (x_{S q})=\gamma_{q} \xi_{f q} f'_{l}(x_{F q})  \label{ftc}, \\
f_{l} (x_{S q})= f_{l}(x_{F q}) + \eta_{q} \,
 \gamma_{bq}\xi_{fq} f'_{l}(x_{F q})  \label{ft},
\end{eqnarray}
with $l=t_{y},t_{z}$. Note that $\eta_{1}=+1$ and $\eta_{2}=-1$. Similarly, we get for the singlet amplitude
\begin{eqnarray}
\xi_{s} f_{s}^{\prime}(x_{S q})=\gamma_{q} \xi_{fq}  f'_{s} (x_{F q}) \label{fsp}, \\
f_{s}(x_{S q})= f_{s}(x_{F q}) + \eta_{q}
 \gamma_{bq} \xi_{fq} f'_{s}(x_{F q}) \label{fss}.
\end{eqnarray}
Here, $x_{F1}$ and $x_{S1}$ are the coordinates on the two sides of
the F$_{1}$/S interface at $x_1=0$, while $x_{F2}$ and $x_{S2}$ refers
to the S/F$_{2}$ interface at $x_2=d_s$. From Eqs.~(\ref{fss}) and
(\ref{ft}) for the first interface (F$_{1}$/S), we obtain the system
\begin{eqnarray}
f_{s}(x_{1})&=&\sum_{\varepsilon=\pm} a_{\varepsilon} {\cal A}_{\varepsilon},
 \label{comp1}\\
c_{y} &=& a_{0} {\cal A}_{0},\\
c_{z} &=& \sum_{\varepsilon=\pm} \varepsilon a_{\varepsilon} {\cal A}_{\varepsilon},
\end{eqnarray}
with the quantity
\begin{eqnarray}
{\cal A}_{j}=\cosh(k_{j 1} d_{f1})+\gamma_{b1} k_{j 1}\xi_{f1} \sinh(k_{j 1} d_{f1}),
\end{eqnarray}
where $j=\pm, 0$.  The matching of the different components with the
conditions~(\ref{fss}) and (\ref{ft}) yield at the second interface
(S/F$_{2}$)
\begin{eqnarray}
f_{s}(x_{2}) &=&\sum_{\varepsilon=\pm} b_{\varepsilon} {\cal B}_{\varepsilon}, \label{comp2}\\
c_{y} \cosh(k_{s} d_{s})+ d_{y} \sinh(k_{s}d_{s}) &=& \sum_{\varepsilon=\pm} \varepsilon b_{\varepsilon} {\cal B}_{\varepsilon} \sin \theta 
-b_{0}{\cal B}_{0} \cos \theta,\\
c_{z} \cosh(k_{s} d_{s})+ d_{z} \sinh(k_{s}d_{s})&=& \sum_{\varepsilon=\pm} \varepsilon b_{\varepsilon} {\cal B}_{\varepsilon}\cos \theta 
+b_{0} {\cal B}_{0} \sin \theta,
\end{eqnarray}
with
\begin{eqnarray}
{\cal B}_{j}=\cosh(k_{j 2} d_{f2})+\gamma_{b2} k_{j 2}\xi_{f2} \sinh(k_{j 2} d_{f2})
\end{eqnarray}
defined in a similar way as the quantity ${\cal A}_{j}$.  Then, the
boundary conditions (\ref{ftc}) yield the system
\begin{eqnarray}
d_{y} &=& a_{0} {\cal C}_{0},\\
d_{z} &=& \sum_{\varepsilon=\pm} \varepsilon a_{\varepsilon} {\cal C}_{\varepsilon},
\end{eqnarray}
for the F$_{1}$/S interface,
and 
\begin{eqnarray}
c_{y} \sinh (k_{s}d_{s})+d_{y} \cosh(k_{s}d_{s}) &=& b_{0} {\cal D}_{0} \cos \theta- \sum_{\varepsilon=\pm} \varepsilon b_{\varepsilon} {\cal D}_{\varepsilon} \sin \theta,
\\
c_{z} \sinh (k_{s}d_{s})+d_{z} \cosh(k_{s}d_{s}) &=& - b_{0} {\cal D}_{0} \sin \theta-\sum_{\varepsilon=\pm} \varepsilon b_{\varepsilon} {\cal D}_{\varepsilon} \cos \theta,
\end{eqnarray}
for the S/F$_{2}$ interface, with
\begin{eqnarray}
{\cal C}_{j} = \gamma_{1} k_{j 1}\xi_{f1} \sinh(k_{j 1}d_{f1})/k_{s} \xi_{s}, \\
{\cal D}_{j} = \gamma_{2} k_{j 2}\xi_{f2} \sinh(k_{j 2}d_{f2})/k_{s} \xi_{s} .
\end{eqnarray}

The next step consists of eliminating the coefficients
$c_{y},c_{z},d_{y},d_{z},a_{0},b_{0}$ from the former equations. We
obtain the system
\begin{eqnarray}
\sum_{\varepsilon} \varepsilon a_{\varepsilon} {\cal E}_{\varepsilon}
&=&
\sum_{\varepsilon} \varepsilon b_{\varepsilon} {\cal G}_{\varepsilon}
\label{syst1},
\\
\sum_{\varepsilon} \varepsilon b_{\varepsilon} {\cal F}_{\varepsilon}
&=&
\sum_{\varepsilon} \varepsilon a_{\varepsilon}
 {\cal H}_{\varepsilon}
\label{syst2},
\end{eqnarray}
where 
\begin{eqnarray}
{\cal E}_{\varepsilon}&=&
K_{0}({\cal A}_{\varepsilon}-{\cal C}_{\varepsilon})\left[\cosh(k_{s}d_{s})-\sinh(k_{s}d_{s})\right] \cos \theta \label{e},
\\
{\cal F}_{\varepsilon}&=&K_{\varepsilon}({\cal B}_{0}-{\cal D}_{0})\sin^{2}\theta +K_{0}({\cal B}_{\varepsilon}-{\cal D}_{\varepsilon})\cos^{2}\theta , \hspace*{1cm}
\\
{\cal G}_{\varepsilon}&=& 
K_{\varepsilon}({\cal B}_{0}+{\cal D}_{0})\sin^{2}\theta +K_{0}({\cal B}_{\varepsilon}+{\cal D}_{\varepsilon})\cos^{2}\theta ,
\\
{\cal H}_{\varepsilon}&=&  
K_{0}({\cal A}_{\varepsilon}+{\cal C}_{\varepsilon})\left[\cosh(k_{s}d_{s})+\sinh(k_{s}d_{s})\right] \cos \theta \label{h},
\end{eqnarray}
with
\begin{eqnarray}
K_{j} =
\left[{\cal B}_{j}{\cal C}_{0}+{\cal D}_{j}{\cal A}_{0}
\right]
 \cosh(k_{s}d_{s}) +
\left[ 
{\cal B}_{j}{\cal A}_{0}+{\cal D}_{j}{\cal C}_{0}
\right]
 \sinh(k_{s}d_{s}). 
\end{eqnarray}
Compiling Eqs.~(\ref{comp1}) and (\ref{comp2}) with
Eqs.~(\ref{syst1})-(\ref{syst2}), we get the expressions for the
amplitudes $a_{\varepsilon}$ and $b_{\varepsilon}$
\begin{eqnarray}
a_{\varepsilon} &=& \frac{\left({\cal B}_{+} {\cal I}_{-,-\varepsilon}+ {\cal B}_{-} {\cal I}_{+,-\varepsilon}\right)f_{s}(x_{1})
+\varepsilon {\cal A}_{-\varepsilon}\left( {\cal F}_{-}{\cal G}_{+}-{\cal F}_{+}{\cal G}_{-}\right)f_{s}(x_{2})}
{{\cal J}}
,
\\
b_{\varepsilon} &=& \frac{\left({\cal A}_{+} {\cal I}_{-\varepsilon,-}+ {\cal A}_{-} {\cal I}_{-\varepsilon,+}\right)f_{s}(x_{2})
+\varepsilon {\cal B}_{-\varepsilon}\left( {\cal E}_{-}{\cal H}_{+}-{\cal E}_{+}{\cal H}_{-}\right)f_{s}(x_{1})}
{{\cal J}}
,
\end{eqnarray}
with 
\begin{eqnarray}
{\cal I}_{\varepsilon, \varepsilon'}&=&{\cal F}_{\varepsilon} {\cal E}_{\varepsilon'}-{\cal G}_{\varepsilon}{\cal H}_{\varepsilon'}
,
\\
{\cal J}&=&{\cal A}_{+}{\cal B}_{+}{\cal I}_{-,-}+{\cal A}_{-}{\cal B}_{-}
{\cal I}_{+,+}
+{\cal A}_{+}{\cal B}_{-}{\cal I}_{+,-}+{\cal A}_{-}{\cal B}_{+}{\cal I}_{-,+}.
\end{eqnarray}

Finally, Eqs. (\ref{fsp}) yield the system
\begin{eqnarray}
\xi_{s} f_{s}^{\prime}(x_{1}) &=& k_{s} \xi_{s} \sum_{\varepsilon} a_{\varepsilon}
{\cal C}_{\varepsilon}
,
\\
\xi_{s} f_{s}^{\prime}(x_{2}) &=& - k_{s} \xi_{s} \sum_{\varepsilon} b_{\varepsilon}
{\cal D}_{\varepsilon},
\end{eqnarray}
which can be rewritten in the form
\begin{eqnarray}
\left(\begin{array}{c}  f_{s}^{\prime}(x_{1})
\\
 f_{s}^{\prime}(x_{2})
\end{array}
\right)
=
k_{s} \left(\begin{array}{cc}
W_{11} & W_{12} \\
W_{21} & W_{22}
\end{array}
\right)
\left(\begin{array}{c}  f_{s}(x_{1})
\\
 f_{s}(x_{2})
\end{array}
\right), \label{bc}
\end{eqnarray}
with
\begin{eqnarray}
W_{11}&=& 
\frac{{\cal C}_{+} \left({\cal B}_{+} {\cal I}_{-,-}+ {\cal B}_{-} {\cal I}_{+,-}\right)
+{\cal C}_{-}\left({\cal B}_{+} {\cal I}_{-,+}+ {\cal B}_{-} {\cal I}_{+,+}\right)}{\cal J},\\
W_{22}&=& -
\frac{{\cal D}_{+} \left({\cal A}_{+} {\cal I}_{-,-}+ {\cal A}_{-} {\cal I}_{-,+}\right)
+{\cal D}_{-}\left({\cal A}_{+} {\cal I}_{+,-}+ {\cal A}_{-} {\cal I}_{+,+}\right)}{\cal J},
\\
W_{12}&=& \frac{\left( {\cal F}_{-}{\cal G}_{+}-{\cal F}_{+}{\cal G}_{-}\right)
\left( {\cal A}_{-}{\cal C}_{+}-{\cal A}_{+}{\cal C}_{-}\right)}{\cal J}, 
\\
W_{21}&=& -\frac{\left( {\cal E}_{-}{\cal H}_{+} - {\cal E}_{+}{\cal H}_{-}\right)
\left( {\cal B}_{-}{\cal D}_{+}-{\cal B}_{+}{\cal D}_{-}\right)}{\cal J}. 
\end{eqnarray}
Using the expressions (\ref{e})-(\ref{h}), one can notice that in fact $W_{12}=-W_{21}$ with
\begin{equation}
W_{12}=
\frac{2 K_{0}^{2} 
\cos^{2} \theta \, 
\left(
{\cal B}_{-}{\cal D}_{+}-{\cal B}_{+}{\cal D}_{-}
\right)
\left(
{\cal A}_{-}{\cal C}_{+}-{\cal A}_{+}{\cal C}_{-}
\right)}{{\cal J}} .
\end{equation}
For an asymmetric trilayer F$_{1}$-S-F$_{2}$, the diagonal
coefficients $W_{11}$ and $W_{22}$ of the matrix $\hat{W}$ differ in
general. In the special case of a symmetric trilayer
F$_{1}$-S-F$_{1}$, we have $W_{11}=W_{22}$.\\

\end{widetext}

\section{Derivation of $\hat W$ for the Pentalayer}\label{App_pentalayer}

Due to the symmetry of the geometry, we need to determine the
components of the anomalous Green function $f$ only in half of the
pentalayer, e.g. in the domain $x>0$. The problem is mapped back onto
the asymmetrical F$_{1}$-S-F$_{2}$ trilayer problem previously
considered in Appendix \ref{App_trilayer}.  Because we have chosen a
different origin for the system of coordinates, the F$_{1}$/S and
S/F$_{2}$ interfaces are now located at the positions $x_{1}=d_{f1}$
and $x_{2}=d_{s}+d_{f1}$.  Due to the shift in coordinates, we have
used the expressions (\ref{expS}) in the S layer and (\ref{expF2}) in
the F$_{2}$ layer with $x$ replaced by $x-d_{f1}$.

For the rotation type 1, the spatial dependences of the singlet and triplet components of
$f$ in the left F$_{1}$ layer are in the 0-junction
case the same  as in Eq. (\ref{expF10j})
after the shift of coordinate.  In the $\pi$-junction case, the
boundary conditions at the (fictitious) outer surface $x=0$ have
changed, and the spatial dependences in F$_{1}$ are now given by:
\begin{equation}
\left( \begin{array}{c} f_{s} \\{\bf f}_{t}
\end{array}
\right)
= \sum_{\varepsilon=\pm} a_{\varepsilon} \sinh\left[k_{\varepsilon \, 1} x\right]
\left(\begin{array}{c} 1\\
\varepsilon \hat{{\bf z}}
\end{array}
\right)
+ a_{0} \sinh\left[k_{0 \, 1} x \right] \left(\begin{array}{c} 0\\
\hat{{\bf x}}
\end{array}
\right). \nonumber
\end{equation}
The new boundary conditions (\ref{bound10}) and (\ref{bound1pi}) at $x=0$ do not affect the definition of
the former quantities ${\cal B}_{j}$ and ${\cal D}_{j}$.  On the other
hand, changes occur in the definition of the quantities ${\cal A}_{j}$
and ${\cal C}_{j}$ (where $j =\varepsilon$ or $0$). In the 0-junction
case, the coefficients ${\cal A}_{j}$ and ${\cal C}_{j}$ remain
unchanged, while in the $\pi$-junction case they are defined as
\begin{eqnarray}
{\cal A}_{j}&=&\sinh(k_{j 1} d_{f1})+\gamma_{b1} k_{j 1}\xi_{f1} \cosh(k_{j 1} d_{f1}), \\
{\cal C}_{j} &=& \gamma_{1} k_{j 1}\xi_{f1} \cosh(k_{j 1}d_{f1})/k_{s} \xi_{s}.
\end{eqnarray}

For the rotation type 2, the components of $f$ in F$_{1}$ have a different spatial dependence as a result of the conditions (\ref{bound20}) or (\ref{bound2pi}). They are expressed as  
\begin{equation}
\left( \begin{array}{c} f_{s} \\{\bf f}_{t}
\end{array}
\right)
= \sum_{\varepsilon=\pm} a_{\varepsilon} \cosh\left[k_{\varepsilon \, 1} x\right]
\left(\begin{array}{c} 1\\
\varepsilon \hat{{\bf z}}
\end{array}
\right)
+ a_{0} \sinh\left[k_{0 \, 1} x \right] \left(\begin{array}{c} 0\\
\hat{{\bf x}}
\end{array}
\right) \nonumber
\end{equation}
in the 0-junction case, and
\begin{equation}
\left( \begin{array}{c} f_{s} \\{\bf f}_{t}
\end{array}
\right)
= \sum_{\varepsilon=\pm} a_{\varepsilon} \sinh\left[k_{\varepsilon \, 1} x\right]
\left(\begin{array}{c} 1\\
\varepsilon \hat{{\bf z}}
\end{array}
\right)
+ a_{0} \cosh\left[k_{0 \, 1} x \right] \left(\begin{array}{c} 0\\
\hat{{\bf x}}
\end{array}
\right) \nonumber
\end{equation}
in the $\pi$-junction case. As for rotation type 1, changes occur in the definition of the quantities ${\cal A}_{j}$ and ${\cal C}_{j}$ (where $j=\varepsilon$ or 0) for the rotation type 2. In the 0-junction case, the coefficients ${\cal A}_{\varepsilon}$ and ${\cal C}_{\varepsilon}$ have the same expression as in Appendix \ref{App_trilayer}, while ${\cal A}_{0}$ and ${\cal C}_{0}$ are now given by
\begin{eqnarray}
{\cal A}_{0}&=&\sinh(k_{0 1} d_{f1})+\gamma_{b1} k_{j 1}\xi_{f1} \cosh(k_{0 1} d_{f1}), \\
{\cal C}_{0} &=& \gamma_{1} k_{0 1}\xi_{f1} \cosh(k_{0 1}d_{f1})/k_{s} \xi_{s}.
\end{eqnarray}
In the $\pi$-junction case, the quantities ${\cal A}_{0}$ and ${\cal C}_{0}$ are defined in the same way as in Appendix \ref{App_trilayer}, while ${\cal A}_{\varepsilon}$ and ${\cal C}_{\varepsilon}$ are written as 
\begin{eqnarray}
{\cal A}_{\varepsilon}&=&\sinh(k_{\varepsilon 1} d_{f1})+\gamma_{b1} k_{\varepsilon 1}\xi_{f1} \cosh(k_{0 1} d_{f1}), \\
{\cal C}_{\varepsilon} &=& \gamma_{1} k_{\varepsilon 1}\xi_{f1} \cosh(k_{\varepsilon 1}d_{f1})/k_{s} \xi_{s}.
\end{eqnarray}

Except for these modifications in the definition of the quantities
${\cal A}$ and ${\cal C}$, the remaining calculations are exactly the
same as in the asymmetric trilayer geometry and we can use the final
expression derived in Appendix \ref{App_trilayer} for the matrix
$\hat{W}$ in the symmetric pentalayer structure.

\section{Derivation of $G(\varepsilon_n,x,y)$}\label{App_G}

In analogy with Ref.~\onlinecite{Fom2002}, Eq.~(\ref{G}) is solved by
making the following ansatz
\begin{equation}
G(x,y) = \left\{
\begin{array}{ll}
L_c(y) X_1(x) + L_s(y) X_2(x), & x<y,\\
R_c(y) Y_1(x) + R_s(y) Y_2(x), & y<x,
\end{array}
\right. \label{G_ansatz}
\end{equation}
where we introduced the notation
\begin{eqnarray}
X_1(x) &=& \cosh(k_s x),\\
X_2(x) &=& \sinh(k_s x),\\
Y_1(x) &=& \cosh\left(k_s [x-d_s]\right),\\
Y_2(x) &=& \sinh\left(k_s [x-d_s]\right).
\end{eqnarray}
The coefficients $L_{c}$, $L_{s}$, $R_{c}$ and $R_{s}$ depend on the
location $y$ of the source term in Eq.~(\ref{G}). The source is taken
into account by the conditions
\begin{equation}
\left. G(x,y) \right|_{x=y^{+}} = \left. G(x,y) \right|_{x=y^{-}}, \label{xy1}
\end{equation}
and
\begin{equation}
\left. \partial_{x}G(x,y) \right|_{x=y^{+}}-\left. \partial_{x} G(x,y) \right|_{x=y^{-}} = -k_{s}^{2}/\varepsilon_{n}, \label{xy2}
\end{equation}
where $y^+$ and $y^-$ denote the limits $y\rightarrow x$ from above
and below, respectively. Eqs.~(\ref{xy1})-(\ref{xy2}) give two
relations between the coefficients in Eq.~(\ref{G_ansatz}). Two
additional relations are provided by the boundary conditions at the
edges of the superconductor, which read
\begin{eqnarray}
\left(
\begin{array}{c}
\left.\partial_{x} G(x,y)\right|_{x=0} \\
\left.\partial_{x} G(x,y)\right|_{x=d_s}
\end{array}
\right)
=k_{s} \hat{W}
\left(
\begin{array}{c}
G(0,y) \\
G(d_{s},y)
\end{array}
\right).
\label{Gbound}
\end{eqnarray}
These conditions are consistent with the boundary conditions
(\ref{bound}) obeyed by the singlet amplitude $f_{s}(x)$. Compiling
Eqs.~(\ref{G_ansatz})-(\ref{Gbound}), we obtain the coefficients
%
%
\begin{eqnarray*}
L_c(y) &=& \frac{k_s}{\varepsilon_n{\cal L}} \left[ Y_1(y) + W_{22}Y_2(y) - W_{12}X_2(y) \right],\\
R_c(y) &=& \frac{k_s}{\varepsilon_n{\cal L}} \left[ X_1(y) - W_{21}Y_2(y) + W_{11}X_2(y) \right],\\
L_s(y) &=& \frac{k_s}{\varepsilon_n{\cal L}} \left[ W_{11}Y_1(y) + W_{12}X_1(y) + \mathrm{det}(\hat W)Y_2(y) \right],\\
R_s(y) &=& \frac{k_s}{\varepsilon_n{\cal L}} \left[ W_{21}Y_1(y) + W_{22}X_1(y) + \mathrm{det}(\hat W)X_2(y) \right],
\end{eqnarray*}
where
\begin{eqnarray*}
{\cal L} &=& W_{12}-W_{21}+(W_{11}-W_{22}) \cosh(k_{s}d_{s}) \nonumber \\
&&+\left[1-\mathrm{det}(\hat{W})\right]\sinh(k_{s}d_{s}).
\end{eqnarray*}
We note that the dependence on the Matsubara frequency $\varepsilon_n$
enters through $k_s$ and the four elements $W_{11}$, $W_{22}$,
$W_{12}$, and $W_{22}$ of the $2\times 2$ matrix $\hat W$ in the
boundary condition.

\section{Derivation of $m_{lp}$ in Eq.~(\ref{eigenp})}\label{App_fourier}

We insert the expansion~(\ref{fourier}) into Eq.~(\ref{fsfromdelta}),
use the expression for $G(\varepsilon_n,x,y)$ derived in Appendix
\ref{App_G}, and perform the integration over the spatial coordinate
$y$. We obtain the singlet amplitude $f_{s}(\varepsilon_n,x)$ in terms
of the Fourier coefficients $\Delta_{p}$
\begin{widetext}
\begin{eqnarray}
f_{s}(\varepsilon_n,x) &=& \frac{\pi}{\varepsilon_{n} {\cal L}}
\sum_{p=0}^{\infty} \Delta_{p}\beta_{p} \left\{ {\cal L} \cos\left(\frac{\pi p x}{d_{s}}\right)
+\left[W_{21}+(-1)^{p} W_{22}\right]X_1(x)
-\left[W_{11}+(-1)^{p} W_{12}\right]Y_1(x) \right.\nonumber\\
&&\hspace{2.4cm}\left.+\,\mathrm{det}(\hat{W})\left[(-1)^{p}X_2(x) - Y_2(x)\right] \frac{}{}\right\},\label{fs_fourier}
\end{eqnarray}
%
%
where $\beta_p=1/\left[1+(\pi p/k_s d_s)^2\right]$ and the functions
${\cal L}$, $X_1(x)$, $X_2(x)$, $Y_1(x)$, and $Y_2(x)$ were introduced
in Appendix~\ref{App_G}. We insert this expression in the gap equation
(\ref{self}) and project in Fourier space, i.e. we multiply by
$\cos(\pi lx/d_s)$ and integrate over $x$. As a result, we obtain a
linear system for the Fourier components $\Delta_{p}$, with row $l \geq 0$
given by
\begin{equation}
\sum_{p=0}^{+\infty} m_{lp} \Delta_{p}=0.\label{lin_syst}
\end{equation}
The off-diagonal elements ($l \neq p$) have the form
\begin{equation}
m_{lp} = 4 \pi T \sum_{\varepsilon_n > 0} \frac{1}{\varepsilon_{n}}  \, b_{lp} \, \beta_l\beta_p,
\label{mlp}
\end{equation}
while the diagonal elements ($l=p$) are given by
\begin{equation}
m_{ll} = (1+\delta_{l0})\ln \frac{T}{T_{c0}}
+4\pi T \sum_{\varepsilon_n>0} \frac{1}{\varepsilon_{n}}
\left[ b_{ll} \beta_l^2 + \frac{1}{2}\left( 1-\beta_l \right) \right],
\label{mll}
\end{equation}
where
\begin{equation}
b_{lp}= \frac{
\left[W_{11}-(-1)^{l+p}W_{22}+
(-1)^{p} W_{12}
-(-1)^{l} W_{21}\right]\sinh(k_{s}d_{s})
+\mathrm{det} (\hat{W}) \left\{(-1)^{p}+(-1)^{l}
-[1+(-1)^{l+p}]\cosh(k_{s}d_{s})\right\}
}{k_{s}d_{s}{\cal L}}. \label{blp}
\end{equation}
\end{widetext}
The relation $W_{12}=-W_{21}$ between the off-diagonal elements of
$\hat{W}$ found in Appendix \ref{App_trilayer} implies that the matrix
$\hat{m}$ is actually symmetric, i.e. $m_{lp}=m_{pl}$ (see expressions
(\ref{mlp}) and (\ref{blp})). This property guarantees the existence
of real solutions of the eigenproblem (\ref{eigenp}).

\section{High-Energy Asymptotics}\label{App_HE}

We present and compare the asymptotic high-energy behavior of the
quasiclassical Green's function in the diffusive limit within the
Usadel approximation to the more general case described by the
Eilenberger equation. Since the present discussion is independent of
the presence or absence of a weak exchange field $J\ll\epsilon_f$ in
the system, we leave it out.

\subsection{Diffusive Limit}

The Usadel equation\cite{Ref_Usadel} for arbitrary temperatures (not
necessarily close to $T_c$ as in the rest of the paper) is
\begin{equation}
\left[ i\varepsilon_n \hat\tau_3 - \hat\Delta, \hat g \right] 
+ \frac{D}{\pi}\partial_x\left(\hat g \partial_x \hat g\right) = \hat 0,
\label{Usadel}
\end{equation}
where $\hat g$ is a $4\times 4$ matrix in combined particle-hole and
spin spaces, $\hat\tau_j$ ($j=1$, $2$, $3$) are the Pauli matrices in
particle-hole space, and $\hat\Delta$ is the gap function
($\hat\Delta=(i\sigma_y)\hat\tau_1\Delta$ if $\Delta$ is
real). Eq.~(\ref{Usadel}) is supplemented with a normalization
condition
\begin{equation}
\hat g^2 = -\pi^2\hat 1.
\end{equation}
Further details concerning  the structure of the Green's function with the
present notation can be found in Ref.~\onlinecite{Cha2005b} (see also
Ref.~\onlinecite{Ale85}).

At high energies the order parameter and the derivative term are small,
\begin{eqnarray}
&& \Delta \sim T_{c0} \ll \varepsilon_n,\\
&& D/\xi^2 \sim T_{c0} \ll \varepsilon_n,
\end{eqnarray}
and we expand the Green's function
\begin{equation}
\label{expansion}
\hat g = \hat g^{(0)} + \hat g^{(1)} + \hat g^{(2)} + ...
\end{equation}
where the term $\hat g^{(k)}$ is of order $(T_{c0}/\varepsilon_n)^k$. To
lowest order we have
\begin{eqnarray}
&& \left[ i\varepsilon_n \hat\tau_3, \hat g^{(0)} \right] = 0,\\
&& \left(\hat g^{(0)}\right)^2 = -\pi^2\hat 1,
\end{eqnarray}
with the solution
\begin{equation}
\label{g0}
\hat g^{(0)} = (-i\pi)\mathrm{sgn}(\varepsilon_n) \hat\tau_3.
\end{equation}
In first order we obtain
\begin{eqnarray}
&& \left[ i\varepsilon_n\hat\tau_3, \hat g^{(1)} \right] = \left[\hat\Delta,\hat g^{(0)}\right],\\
&& \hat g^{(0)} \hat g^{(1)} + \hat g^{(1)} \hat g^{(0)} = 0.
\end{eqnarray}
Since $\hat g^{(0)}$ is proportional to $\hat\tau_3$, the second line
can be used to move $\hat g^{(1)}$ to one side of the commutator on
the left-hand-side of the first line. We obtain
\begin{equation}
2i\varepsilon_n\hat\tau_3\hat g^{(1)} = \left[\hat\Delta,\hat g^{(0)}\right].
\end{equation}
The solution is purely off-diagonal in particle-hole space
\begin{equation}
\label{g1}
\hat g^{(1)} = \frac{(-i\pi)}{2i|\varepsilon_n|}\left(\hat\tau_3\hat\Delta\hat\tau_3-\hat\Delta\right)
= \frac{\pi}{|\varepsilon_n|}\hat\Delta.
\end{equation}
In second order we have
\begin{eqnarray}
&& \left[ i\varepsilon_n\hat\tau_3,\hat g^{(2)} \right] = -\frac{D}{\pi}\hat g^{(0)}\partial_x^2\hat g^{(1)}, \\
&& \hat g^{(0)} \hat g^{(2)} + \hat g^{(2)} \hat g^{(0)} + \left(\hat g^{(1)}\right)^2 = 0.
\label{n2}
\end{eqnarray}
After a short calculation, similar to the calculation in first order,
we obtain
\begin{equation}
\label{g2}
\hat g^{(2)} = \frac{(-i\pi)}{2\varepsilon_n|\varepsilon_n|}\hat\tau_3\hat\Delta^2
+ \frac{\pi D}{2\varepsilon_n^2} \partial_x^2\hat\Delta.
\end{equation}
Note, in particular, that there is an off-diagonal term proportional
to $1/\varepsilon_n^2$ for inhomogeneous systems.

The off-diagonal part of the Green's function has according to the
above the asymptotic form
\begin{equation}
f(\varepsilon_n,x) = \frac{\pi\Delta(x)}{|\varepsilon_n|} + \frac{\pi D \partial_x^2\Delta(x)}{2\varepsilon_n^2}
+{\cal O}\left[\left(\frac{T_{c0}}{|\varepsilon_n|}\right)^3\right],
\label{HEf}
\end{equation}
which we now use to discuss the gap equation. The gap equation
\begin{equation}
\Delta(x) = \lambda\,T\sum_{|\varepsilon_n|<\omega_p} f(\varepsilon_n,x),
\end{equation}
contains a log-divergency and it is necessary to introduce a cut-off
$\omega_p$.  But by the well-known procedure (see
e.g. Ref.~\onlinecite{Xu95}), the interaction strength $\lambda$ and
the Matsubara sum cut-off $\omega_p$ can both be eliminated by adding
and subtracting the leading high-energy term in Eq.~(\ref{HEf}). The
gap equation then has the form in Eq.~(\ref{gap}).  The Matsubara sum
converges, with a high-energy asymptotic tail $\propto 1/\varepsilon_n^2$
according to Eq.~(\ref{HEf}), and can be extended to infinity. In
practice, a technical cut-off $\varepsilon_c$ is introduced that
should, however, be high enough that the results of the calculation
are cut-off independent.

\subsection{Arbitrary Mean Free Path}

We compare the above results obtained within the Usadel approximation
with the corresponding high-energy behavior obtained within the
Eilenberger approach. The Eilenberger
equation\cite{Ref_Eilenberger,Ref_LO} reads
\begin{equation}
\left[ i\varepsilon_n \hat\tau_3 - \hat\Delta - \hat \sigma_{imp} , \hat g \right] 
+ i \vfgrad \hat g = \hat 0,
\label{Eilen}
\end{equation}
with impurity self energy $\hat \sigma_{imp}$, 
and where $\vvf $ is the Fermi velocity.
The normalization
condition $\hat g^2=-\pi^2\hat 1 $ holds.
We include non-magnetic 
impurity scattering within the self-consistent $t$-matrix approximation,
for which the impurity self energy is
\begin{equation}
\hat \sigma_{imp}(\vpf )=c\; \hat t(\vpf ,\vpf ), 
\end{equation}
where $c$ is the impurity concentration, and $s$ is a parameter that 
specifies the position
of the momentum on the Fermi surface.
The $t$-matrix is given as the solution of the equation
\begin{eqnarray}
\hat t(\vpf ,\vpfp)&=& \hat u(\vpf ,\vpfp )+ \nonumber \\
&&
\Big\langle \hat u(\vpf ,\vpfpp )\cNf (\vpfpp )\hat g(\vpfpp ) 
\hat t (\vpfpp ,\vpfp ) \Big\rangle_{\vpfpp }, \qquad \quad
\label{Impurity}
\end{eqnarray}
where we have omitted for brevity all variables except the Fermi-momentum.
Here, $\hat u(\vpf, \vpfp )=u(\vpf, \vpfp ) \hat 1$ is the impurity scattering potential, and
$\langle ... \rangle_{\vpfp} $ denotes a Fermi surface average over
$\vpfp$ .

We expand $\hat g$ as in Eq.~(\ref{expansion}). The zeroth order 
term for the Green function is given analogously to the discussion 
for the diffusive limit by
\begin{eqnarray}
&& \hat g^{(0)} = (-i\pi)\mathrm{sgn}(\varepsilon_n)\hat\tau_3.
\end{eqnarray}
For the higher orders we need to expand the impurity $t$-matrix in the
parameter $(T_{c0}/\varepsilon_n)$,
\begin{eqnarray}
\label{expansions}
\hat t &=& \hat t^{(0)} + \hat t^{(1)} + 
\hat t^{(2)} + ...,
\end{eqnarray}
and similarly for the impurity self energy.
Introducing the operator 
\begin{equation}
\hat D(\vpf,\vpfp )=  \delta(\vpf-\vpfp )\hat 1 - u(\vpf,\vpfp) \cNf (\vpfp ) 
\hat g^{(0)} 
\end{equation}
the $t$-matrix equation for the zeroth order term $\hat t^{(0)}$ 
takes the form
\begin{equation}
\left\langle \hat D(\vpf,\vpfpp ) \hat t^{(0)}(\vpfpp, \vpfp )\right\rangle_{\vpfpp}=  
\hat u(\vpf,\vpfp).
\end{equation}
With the inverse operator $\hat D^{-1}$ defined by
\begin{equation}
\left\langle \hat D^{-1}(\vpf,\vpfpp ) \hat D(\vpfpp, \vpfp )\right\rangle_{\vpfpp}=  
\delta(\vpf-\vpfp) \hat 1
\end{equation}
the formal solutions are given by
\begin{eqnarray}
\label{t0}
 \hat t^{(0)}(\vpf,\vpfp) &=& 
 \left\langle \hat D^{-1}(\vpf,\vpfpp ) \hat u (\vpfpp,\vpfp)
\right\rangle_{\vpfpp}\\
 \hat t^{(1)}(\vpf,\vpfp) &=& \left\langle \hat t^{(0)}(\vpf,\vpfpp) \cNf (\vpfpp)\hat g^{(1)}(\vpfpp) \hat t^{(0)}(\vpfpp,\vpfp)\right\rangle_{\vpfpp}
 \nonumber
\label{t1}.
\end{eqnarray}
From Eq.~(\ref{t0}) we obtain,
\begin{equation}
[\hat \sigma_{imp}^{(0)},\hat g^{(0)}] = \hat 0,
\end{equation}
as a result of $[\hat u,\hat \tau_3]=\hat 0$. Consequently, the first order
term for $\hat g$ is, in complete analogy to the discussion leading to
Eq.~(\ref{g1}), given by
\begin{eqnarray}
&& \hat g^{(1)} = \frac{\pi}{|\varepsilon_n|}\hat\Delta.
\end{eqnarray}
Finally, for the second order term $\hat g^{(2)}$, we have
\begin{equation}
[i\varepsilon_n \hat \tau_3,\hat g^{(2)}] = [\hat \sigma_{imp}^{(1)},\hat g^{(0)}]
+[\hat \sigma_{imp}^{(0)},\hat g^{(1)}]
-i\vfgrad \hat g^{(1)}.
\end{equation}
We solve this equation by using the normalization condition, Eq.~(\ref{n2}).
Restricting ourselves to isotropic impurity scattering,
we obtain\cite{eschrigThesis}
%
\begin{eqnarray}
\hat g^{(2)} = \frac{(-i\pi)}{2\varepsilon_n|\varepsilon_n|}&\hat\tau_3 &
\Big( i\vfgrad\hat\Delta - \hat\Delta^2 + \Big. \nonumber \\
&&
\Big. \frac{i\; \mbox{sgn}(\varepsilon_n)}{\tau }
\hat \tau_3 \left\{ \hat \Delta -\langle \hat \Delta \rangle_{\vpf } \right\}
\Big),  
\label{ballistic_O2}
\end{eqnarray}
where 
the inverse scattering time is defined as,
\begin{eqnarray}
\frac{1}{\tau }=2\pi c \cNf \frac{u^2}{1+\pi^2 \cNf^2 u^2}.
\end{eqnarray}
For an isotropic 
($s$-wave) superconducting order parameter the last term in 
Eq.~(\ref{ballistic_O2}) vanishes.
In this case, the second order
high-energy contribution from Eq.~(\ref{ballistic_O2}) is odd in
frequency, and it drops out of the Matsubara
sum. The leading order contribution comes in third order\cite{eschrigThesis}
and the high-energy tail of the Matsubara sum is $\propto
1/|\varepsilon_n|^3$.
This means that the technical cut-off $\varepsilon_c$ can be chosen much
smaller than in the diffusive limit within the Usadel approximation.

The different high-energy asymptotics within the Eilenberger and
Usadel approaches are due to the diffusive approximation employed by
Usadel: the impurity self-energy, i.e. the inverse scattering time
$1/\tau$, is at the outset assumed to be the largest energy scale in
the problem. The high-energy tail is different depending on the order
in which the limits $\tau\rightarrow 0$ and
$\varepsilon_c\rightarrow\infty$ are taken.

\section{Analytic Summation of the High-Energy Tail in the Fourier Series Approach}\label{App_HEfourier}

At high energies $\varepsilon_n\gg T_{c0}$ and $J$, the matrix $\hat W$ has
a simple energy dependence that we exploit to sum the Matsubara sum to
infinity. That is, we write
\begin{equation}
m_{lp} = \bar m_{lp} + {\cal R}_{lp},
\end{equation}
where $\bar m_{lp}$ includes terms in the sum in
Eqs.~(\ref{mlp})-(\ref{mll}) up to a technical cut-off $\varepsilon_c$
while the rest term ${\cal R}_{lp}$ is the sum from $\varepsilon_c$ to
infinity computed analytically below.

At high energies $W_{12}=- W_{21}\approx 0$, while
\begin{eqnarray}
W_{11} &\approx& \frac{\gamma_1}{1+\gamma_{b1}\lambda},\\
W_{22} &\approx& -\frac{\gamma_2}{1+\gamma_{b2}\lambda},
\end{eqnarray}
where $\lambda^2=\varepsilon_n/\pi T_{c0}$. These relations hold for both
the trilayer and the pentalayer, which reflects the fact that the
theory becomes local at high energies (see the effective boundary condition (\ref{bound})). The key function of the Fourier
method then has the form
\begin{equation}
b_{lp} = \frac{\xi_S}{d_S}\frac{2}{\lambda}\frac{c_1+c_2\lambda}{c_3+c_4\lambda+c_5\lambda^2},
\end{equation}
where
\begin{eqnarray}
c_1 &=& \gamma_1+(-1)^{l+p}\gamma_2+\gamma_1\gamma_2\left[1+(-1)^{l+p}\right], \label{c1}\\
c_2 &=& \gamma_1\gamma_{b2}+(-1)^{l+p}\gamma_2\gamma_{b1},\\
c_3 &=& 1+\gamma_1+\gamma_2+\gamma_1\gamma_2,\\
c_4 &=& \gamma_1\gamma_{b2}+\gamma_2\gamma_{b1}+\gamma_{b1}+\gamma_{b2},\\
c_5 &=& \gamma_{b1}\gamma_{b2}. \label{c5}
\end{eqnarray}
For each element of the matrix $m_{lp}$ we can perform the high-energy
Matsubara sum by integration. We get
\begin{eqnarray}
{\cal R}_{lp} &=& \delta_{lp} \frac{1}{\pi} \ln\left(1+\frac{p^2}{\tilde d_s^2}\frac{T_{c0}}{\varepsilon_c}\right) +
\frac{2}{\pi^{2}}\frac{1}{\tilde{d}_S}I_{lp},\label{Rlp}\\
I_{lp} &=& \int_{\frac{\varepsilon_c}{T_{c0}}}^{\infty}\frac{c_1\sqrt{x}+c_2x}
{\left(x+\frac{l^2}{\tilde d_s^2}\right)\left(x+\frac{p^2}{\tilde d_s^2}\right)(c_3+c_4\sqrt{x}+c_5x)} dx,\nonumber
\end{eqnarray}
where we used the short hand notation $\tilde d_s=d_s/\pi\xi_S$. Note 
that Eq. (\ref{Rlp}) is independent of the temperature $T$ and only depends on the parameters in Eqs. (\ref{c1})-(\ref{c5}), on $d_{s}$ and on the cut-off $\varepsilon_{c}$.



\begin{thebibliography}{99}


\bibitem{Buz1990}
A.I. Buzdin and M. Yu. Kupriyanov, JETP Lett. \textbf{52}, 487 (1990).

\bibitem{Rad1991}
Z. Radovic, M. Ledvij, L. Dobrosavljevic-Grujic, A.I. Buzdin, and J.R. Clem, Phys. Rev. B \textbf{44}, 759 (1991).

\bibitem{Buz1992}
A.I. Buzdin, B. Bujicic, and M.Yu. Kupriyanov, Sov. Phys. JETP \textbf{74}, 124 (1992).

\bibitem{Dem1997}
E.A. Demler, G.B. Arnold, and M.R. Beasley, Phys. Rev. B \textbf{55}, 15174 (1997).

\bibitem{Tag1999}
L.R. Tagirov, Phys. Rev. Lett. \textbf{83}, 2058 (1999).

\bibitem{Buz1999}
A.I. Buzdin, A.V. Vedyayev, and N.V. Ryzhanova, Europhys. Lett. \textbf{48}, 686 (1999).

\bibitem{Bal2001}
I. Baladi\'{e}, A. Buzdin, N. Ryzhanova, and A. Vedyayev, Phys. Rev. B \textbf{63}, 054518 (2001).


\bibitem{Fom2002}
Ya.V. Fominov, N.M. Chtchelkatchev, and A.A. Golubov, JEPT Lett. \textbf{74}, 96 (2001); Phys. Rev. B \textbf{66}, 014507 (2002).

\bibitem{Bal2003}
I. Baladi\'{e} and A. Buzdin, Phys. Rev. B \textbf{67}, 014523 (2003).


\bibitem{Fom2003}
Y.V. Fominov, A.A. Golubov, and M.Y. Kupriyanov, JETP Letters \textbf{77}, 510 (2003).

\bibitem{You2004}
C.-Y. You, Ya.B. Bazaliy, J.Y. Gu, S.-J. Oh, L.M. Litvak, and S.D. Bader, Phys. Rev. B \textbf{70}, 014505 (2004).



\bibitem{Jia1995}
J.S. Jiang, D. Davidovi\'{c}, D. H. Reich, and C.L. Chien, Phys. Rev. Lett. \textbf{74}, 314 (1995).

\bibitem{Muh1996}
Th. M\"{u}hge, N.N. Garif'yanov, Yu. V. Goryunov, G.G. Khaliullin, L.R. Tagirov, K. Westerholt, I.A. Garifullin, and H. Zabel, Phys. Rev. Lett. \textbf{77}, 1857 (1996).


\bibitem{Aar1997}
J. Aarts, J.M.E. Geers, E. Br\"{u}ck, A.A. Golubov, and R. Coehoorn, Phys. Rev. B \textbf{56}, 2779 (1997).


\bibitem{Laz2000}
L. Lazar, K. Westerholt, H. Zabel, L.R. Tagirov, Yu. V. Goryunov, N.N. Garif'yanov, and I.A. Garifullin, Phys. Rev. B \textbf{61}, 3711 (2000).


\bibitem{Gu2002}
J.Y. Gu, C.-Y. You, J.S. Jiang, J. Pearson, Ya. B. Bazaliy, and S.D. Bader, Phys. Rev. Lett. \textbf{89}, 267001 (2002).

\bibitem{Obi2005}
Y. Obi, M. Ikebe, and H. Fujishiro, Phys. Rev. Lett. \textbf{94}, 057008 (2005).

\bibitem{Cir2005}
C. Cirillo, S.L. Prischepa, M. Salvato, C. Attanasio, M. Hesselberth, and J. Aarts, Phys. Rev. B \textbf{72}, 144511 (2005).

\bibitem{Pot2005}
A. Potenza and C. H. Marrows, Phys. Rev. B \textbf{71}, 180503(R) (2005).

\bibitem{Mor2006}
I.C. Moraru, W.P. Pratt, and N.O. Birge, Phys. Rev. Lett. \textbf{96}, 037004 (2006).

\bibitem{Rus2006}
A. Yu. Rusanov, S. Habraken, and J. Aarts, Phys. Rev. B \textbf{73}, 060505(R) (2006).

\bibitem{Rus2004}
A.Yu. Rusanov, M. Hesselberth, J. Aarts, and A.I. Buzdin, Phys. Rev. Lett. \textbf{93}, 057002 (2004).

\bibitem{Buz}
A.I. Buzdin, Rev. Mod. Phys. \textbf{77}, 935 (2005).



\bibitem{Buz1982}
A.I. Buzdin, L.N. Bulaevskii, and S.V. Panyukov, JETP Lett. \textbf{35}, 178 (1982).

\bibitem{Buz1991}
A.I. Buzdin and M. Yu. Kupriyanov, JETP Lett. \textbf{53}, 321 (1991).


\bibitem{Rya2001}
V.V. Ryazanov, V.A. Oboznov, A. Yu. Rusanov, A.V. Veretennikov, A.A. Golubov, and J. Aarts, Phys. Rev. Lett. \textbf{86}, 2427 (2001)

\bibitem{Sel2004}
H. Sellier, C. Baraduc, F. Lefloch, and R. Calemczuk, Phys. Rev. Lett. \textbf{92}, 257005 (2004).

\bibitem{Fro2004}
S.M. Frolov, D.J. Van Harlingen, V.A. Oboznov, V.V. Bolginov, and V.V. Ryazanov, Phys. Rev. B \textbf{70}, 144505 (2004).



\bibitem{Kon2002}
T. Kontos, M. Aprili, J. Lesueur, F. Gen\^{e}t, B. Stephanidis, and R. Boursier, Phys. Rev. Lett. \textbf{89}, 137007 (2002).

\bibitem{Blu2002}
Y. Blum, A. Tsukernik, M. Karpovski, and A. Palevski, Phys. Rev. Lett. \textbf{89}, 187004 (2002).


\bibitem{Gui2003}
W. Guichard, M. Aprili, O. Bourgeois, T. Kontos, J. Lesueur, and P. Gandit, Phys. Rev. Lett. \textbf{90}, 167001 (2003).

\bibitem{Shelukin}
V. Shelukin, A. Tsukernik, M. Karpovski, Y. Blum, K.B. Efetov, A.F. Volkov, T. Champel, M. Eschrig, T. L\"{o}fwander, G. Sch\"{o}n, and A. Palevski, cond-mat/0512593, accepted for publication in Phys. Rev. B (2006).


\bibitem{Ber2001c}
F.S. Bergeret. A.F. Volkov, and K.B. Efetov, Phys. Rev. Lett. \textbf{86}, 3140 (2001).

\bibitem{Gol2002}
A.A. Golubov, M. Yu. Kupriyanov, and Ya. V. Fominov, JETP Lett. \textbf{75}, 190 (2002).

\bibitem{Bla2004}
Ya. M. Blanter and F.W.J. Hekking, Phys. Rev. B \textbf{69}, 024525 (2004).

\bibitem{Ber2001b}
F.S. Bergeret, A.F. Volkov, and K.B. Efetov, Phys. Rev. B \textbf{64}, 134506 (2001).


\bibitem{Ber2003} F.S Bergeret, A.F. Volkov, and K.B. Efetov, Phys. Rev. B \textbf{68}, 064513 (2003).

\bibitem{Ber2001}
F.S. Bergeret, A.F. Volkov, and K.B. Efetov, Phys. Rev. Lett. \textbf{86}, 4096 (2001).

\bibitem{Vol2003}
A.F. Volkov, F.S. Bergeret, and K.B. Efetov, Phys. Rev. Lett. \textbf{90}, 117006 (2003).

\bibitem{Kad2001}
A. Kadigrobov, R.I. Shekhter, and M. Jonson, Europhys. Lett. \textbf{54}, 394 (2001) ; Low Temp. Phys. \textbf{27}, 760 (2001).

\bibitem{escPRL03}
M. Eschrig, J. Kopu, J.C. Cuevas, and G. Sch{\"o}n, Phys. Rev. Lett. {\bf 90}, 137003 (2003).

\bibitem{Cha2005}
T. Champel and M. Eschrig, Phys. Rev. B \textbf{71}, 220506(R) (2005).

\bibitem{Cha2005b}
T. Champel and M. Eschrig, Phys. Rev. B \textbf{72}, 054523 (2005).



\bibitem{Lof2005}
T. L\"{o}fwander, T. Champel, J. Durst, and M. Eschrig, Phys. Rev. Lett. \textbf{95}, 187003 (2005).

\bibitem{Ber2005}
F.S. Bergeret, A.F. Volkov, and K.B. Efetov, Rev. Mod. Phys. \textbf{1321} (2005).

\bibitem{Ref_Usadel}
K.D. Usadel, Phys. Rev. Lett. \textbf{25}, 507 (1970).

\bibitem{Kup}
M.Yu. Kupriyanov and V.F. Lukichev, Sov. Phys. JETP \textbf{67}, 1163 (1988).


\bibitem{FFLO1}
P. Fulde and R.A. Ferrell, Phys. Rev. {\bf 135}, A550 (1964).

\bibitem{FFLO2}
A.J. Larkin and Y.N. Ovchinnikov, Zh. Eksp. Teor. Fiz. {\bf 47}, 1136 (1964) [Sov. Phys. JETP {\bf 20}, 762 (1965)].

\bibitem{He-3-FFLO1}
A.B. Vorontsov and J.A. Sauls, Phys Rev B, {\bf 68}, 064508 (2003).

\bibitem{He-3-FFLO2}
A.B. Vorontsov and J.A. Sauls, cond-mat/0601565.

\bibitem{Ale85}
J.A.X Alexander, T.P. Orlando, D. Rainer, and P.M. Tedrow, Phys. Rev. B {\bf 31}, 5811 (1985).

\bibitem{Xu95}
D. Xu, S.-K. Yip, and J.A. Sauls, Phys. Rev. B {\bf 51}, 16233 (1995).

\bibitem{Ref_Eilenberger}
G. Eilenberger, Z. Phys. \textbf{214}, 195 (1968).

\bibitem{Ref_LO}
A.I. Larkin and Y.N. Ovchinnikov, Zh. Eksp. Teor. Fiz. \textbf{55}, 2262 (1968) [Sov. Phys. JETP \textbf{28}, 1200 (1969)].


\bibitem{eschrigThesis}
M. Eschrig, Ph.D. Thesis, Bayreuth University (1997).



\end{thebibliography}
\end{document}